\documentclass{aa}  
\usepackage{graphicx}
\usepackage{subfigure}
\usepackage{multirow}
\usepackage{amssymb}
\usepackage{natbib}
\usepackage{txfonts}
\usepackage{rotating}
\usepackage[bookmarks=false,colorlinks=true, citecolor=blue, backref=true]{hyperref}
\begin{document}
   \title{Observations and modeling of the dust emission from the H$_2$-bright galaxy-wide shock in Stephan's Quintet}


   \author{P. Guillard\inst{1, 2}
          \and
          F. Boulanger\inst{1}
          \and
          M. E. Cluver\inst{2}
		  \and
	      P. N. Appleton\inst{3}
         \and
		  G. Pineau des For\^ets\inst{1,4}
         \and
		  P. Ogle\inst{2}
          }

    \authorrunning{P.~Guillard et al.}
    \titlerunning{Dust emission in the Stephan's Quintet galaxy-wide shock}
    \offprints{P.~Guillard, \email{pierre.guillard@ias.u-psud.fr}}
  
   \institute{Institut d'Astrophysique Spatiale (IAS), UMR 8617, CNRS, Universit\'e
    Paris-Sud 11, B\^atiment 121, 91405 Orsay Cedex, France  \email{pierre.guillard@ias.u-psud.fr}
    \and \textit{Spitzer} Science Center, IPAC, California Institute of Technology, Mail code 100-22, Pasadena, CA 91125, USA
    \and NASA \emph{Herschel} Science Center (NHSC), IPAC, California Institute of Technology, Mail code 100-22, Pasadena, CA 91125, USA
   \and  LERMA, UMR 8112, CNRS, Observatoire de Paris, 61 Avenue de l'Observatoire, 75014 Paris, France
   }

  \date{Received October 2009 / accepted 28th March 2010 }

 
\abstract
{\textit{Spitzer Space Telescope} observations revealed powerful mid-infrared (mid-IR) H$_{2}$ rotational line emission from the Stephan's Quintet (SQ) X-ray~emitting large~scale~shock ($\sim15\times35\,$kpc$^2$) associated~with~a~collision~between~two galaxies. Because  H$_2$ forms on dust grains, the presence of H$_2$ is physically linked to the survival of dust, and we expect some dust emission to come from the molecular gas.
}
{
To test this interpretation, IR~observations and dust modeling are used to identify and characterize the thermal dust emission from the shocked molecular gas.
}
   {The spatial distribution of the IR emission allows us to isolate the faint PAH and dust continuum emission 
associated with the molecular gas in the SQ shock. We model the spectral energy distribution (SED) of this emission, 
and fit it to \textit{Spitzer} observations.
The radiation field is determined with \textit{GALEX} UV, \textit{HST}~$V$-band, and ground-based near-IR observations. We  consider two limiting cases for the structure of the H$_2$ gas. It is either diffuse, penetrated by UV radiation, or fragmented into clouds optically thick to UV. 
}
   {
Faint PAH~and~dust continuum~emission~are~detected in the SQ shock, outside star-forming regions. The $12/24\,\mu$m flux ratio in the shock is remarkably close to that of the diffuse Galactic interstellar medium, leading to a Galactic PAH/VSG abundance~ratio. However, the properties of the PAH emission spectrum in the shock differ from that of the Galaxy, which may suggest an enhanced fraction of large and neutrals PAHs. In both models (diffuse or clumpy H$_2$ gas), the IR~SED is consistent with the expected emission from dust associated with the warm ($>150\,$K) H$_2$ gas, heated by a UV radiation field of intensity comparable to that of the solar neighborhood. This is in agreement with \textit{GALEX UV} observations that show that the intensity  of the radiation field in the shock is $G_{\rm UV} = 1.4 \pm 0.2$ [Habing units].
}
{
 The presence of PAHs and dust grains in the high-speed ($\sim\!1000\,$km~s$^{-1}$) galaxy collision suggests that dust survives. We propose~that~the dust that survived destruction was in pre-shock gas 
at densites larger than a few $0.1\,$cm$^{-3}$, which was not shocked at velocities larger than $\sim \!200\,$km~s$^{-1}$. Our model assumes a Galactic dust-to-gas mass~ratio and size~distribution, and present data do not allow us~to identify any~significant deviations of the abundances and size distribution of dust grains from that of the Galaxy. Our model calculations show that far-IR \textit{Herschel} observations will help constraining the structure of the molecular gas, and the dust size distribution, and thereby to look for signatures of dust processing in the SQ shock.
}

 
\maketitle

\section{Introduction}

Stephan's Quintet (Hickson Compact Group \object{HCG92}, Arp 319, hereafter SQ) 
is an extensively studied compact group of four interacting galaxies that have a complex dynamical history \citep[e.g.][]{Allen1980, Moles1997}. 
A remarkable feature of SQ is that a giant ($\approx 15 \times 35$~kpc) shock is created by an intruding galaxy, \object{NGC 7318b} (Sbc pec), colliding into \object{NGC 7319}'s tidal tail, at a relative velocity of $\sim 1\,000$~km~s$^{-1}$. 
Evidence for a group-wide shock comes from observations of an extended X-ray ridge containing shock-heated ($\sim 5\times 10^6$~K) gas \citep{Pietsch1997, Trinchieri2003, Trinchieri2005, O'Sullivan2009},
strong radio synchrotron emission from the radio emitting plasma
\citep{Allen1972, Sulentic2001, Williams2002, Xu2003} and shocked-gas excitation diagnostics
from optical emission lines \citep{Xu2003}. This extended region is denoted \textit{``ridge''} or simply \textit{``SQ shock''} in this paper. 

Observations with the infrared spectrograph \citep[\textit{IRS},][]{Houck2004} onboard the \textit{Spitzer Space Telescope} have revealed a powerful mid-infrared (mid-IR) $\rm H_2$ rotational line emission from warm ($\sim 10^{2} - 10^{3}\,$K) molecular gas in the SQ shock
\citep{Appleton2006, Cluver2010}. 
The H$_2$ emission is extended not only over the whole ridge, but in several other structures, including an extension towards the Seyfert  galaxy NGC~7319,  and the  intergalactic starburst SQ-A \citep{Xu1999}. The latter structure, beyond the northern tip of the ridge, has been shown to contain significant CO-emitting gas \citep{Gao2000, Smith2001a, Lisenfeld2002}. 
To explain the H$_2$ emission from the SQ ridge, \citet{Guillard2009} considered the collision of two flows of multiphase dusty gas and proposed a model that quantifies the gas cooling, dust destruction, H$_{2}$ formation and excitation in the postshock medium.
H$_2$ gas can form out of 
gas that is shocked to velocities sufficiently low ($V_{\rm s} < 200$~km~s$^{-1}$) for dust to survive. 
Because H$_2$ molecules form on dust grains \citep[e.g.][]{Cazaux2004}, dust is a key element in this scenario.

\citet{Xu1999, Xu2003} reported detection of diffuse far-infrared (hereafter FIR) emission from the intergalatic medium (hereafter IGM) with the \textit{Infrared Space Observatory (ISO)}, and proposed that the dust emission in the shock region would arise from dust grains that efficiently cool the X-ray emitting plasma via collisions with hot electrons. 
The FIR emission would then trace the structure of the shock, as suggested by \citet{Popescu2000a} for the case of shocks driven into dusty gas that is accreting onto clusters of galaxies.
However, the poor spatial resolution of these observations makes it difficult to separate the dust emission associated with star formation (in the neighborhood galaxies, or SQ-A) from that really associated with the shock. 
In addition, \citet{Guillard2009} show that the dust contribution to the cooling of the hot ($\sim 5 \times 10^{6}$~K) plasma  is expected to be low, because of efficient thermal sputtering of the grains. If we assume that the age of the galaxy collision is $\sim 5 \times 10^{6}$~yrs, grains  smaller than $0.1\,\mu$m in radius must have been destroyed. However, the plasma could still be dusty if, before the shock, a significant fraction of the dust mass was  in larger grains and/or if dust destruction is balanced by mass exchange between the cold and the hot gas phases \citep{Guillard2009}. 

The discovery of bright H$_2$ emission in the ridge set new perspectives on the origin of the dust emission in the SQ ridge. In the context of our model for the H$_2$ formation in the SQ shock, we expect some dust emission from the molecular gas. We use \textit{Spitzer} observations and a model of the dust emission to test this expectation.
\textit{Spitzer} observations show that the bright Polycyclic Aromatic Hydrocarbons (henceforth PAHs) and mid-IR continuum emitting regions are spatially correlated with UV emission and associated with star-forming regions mainly associated with the individual sources in the group (in particular the spiral arm of the intruder galaxy NGC~7318b) and with SQ-A \citep{Cluver2010, Natale2010}. 
These IR-bright regions do not correlate with the radio, X-ray, or H$_2$ line emission, which trace the shock structure.
In this paper we focus on the fainter dust emission from the the SQ ridge itself, outside these star-forming regions. 
The spatial distribution of the mid-IR emission allows us to isolate the dust emission from the shock.
UV, visible and near-IR observations determine the spectral energy distribution (hereafter SED) of the radiation ﬁeld 
used as an input to the dust model. We consider two limiting cases for the structure of the molecular gas, either diffuse, or fragmented into  clouds that are optically thick to UV light. An updated version of the \citet{Desert1990} model is used to compute the dust emission from molecular gas in these two cases, and fit the model results to the observed IR SED in the SQ shock. The Galactic dust size distribution is taken as a reference. 



This paper is organized as follows. 
Section~\ref{sec:UVIRobservations} presents the new IR \textit{Spitzer} observations of SQ, and the UV, optical and near-IR ancillary data we use in this paper. The method used to perform photometry within the SQ shock region and the results are described in sect.~\ref{sec:photometry}.
Section~\ref{sec:dustemission} presents the \textit{Spitzer} imaging and spectroscopy results pertaining to the dust emission in the shock structure, emphasizing the PAH properties in the shock.  
The physical framework and inputs of the dust modeling are outlined in  sect.~\ref{sec:models}, and the results are discussed and compared to the mid-IR \textit{Spitzer} observations in sect.~\ref{sec:results}. In sect.~\ref{sec:discussion} we discuss the dust processing  in the shock. Then, we  present our conclusions in sect.~\ref{sec:conclusion} and propose new observations to constrain the physical structure of the molecular gas in the shock.

In this paper we assume the distance to the SQ group to be 94~Mpc (with a Hubble
constant of 70 km~s$^{-1}$~Mpc$^{-2}$) and a systemic velocity for the
group as a whole of $6\,600$~km~s$^{-1}$. At this distance, $10\ \rm arcsec=4.56\,$kpc.

\section{Observations of the Stephan's Quintet shock}
\label{sec:UVIRobservations}

In the following paragraphs we present the new mid-IR (sect.~\ref{subsec:midIRimaging}) observations of SQ, and the ancillary UV (sect.~\ref{UVimaging}), optical (sect.~\ref{subsec:HST_imaging}), and near-IR (sect.~\ref{subsec:nearIRimaging}) data, respectively. 

\subsection{\textit{Spitzer} IR imaging and spectroscopy}
\label{subsec:midIRimaging}

Stephan's Quintet has been imaged with the \textit{InfraRed Array Camera} \citep[\textit{IRAC},][]{Fazio2004} at $3.6$, $4.5$, $5.8$, $8\,\mu$m, with the \textit{IRS} blue peak-up imager (PUI) at $16\,\mu$m, and with the   \textit{Multiband Imaging Photometer for Spitzer} \citep[\textit{MIPS},][]{Rieke2004} at $24$ and $70\,\mu$m. The $70\,\mu$m image has been reported in \citet{Xu2008}. The pixel sizes are $1.8"$, $2.45"$ and $5"$ at $16$,  $24$ and $70\,\mu$m, respectively. 
Except for the $70\,\mu$m image, we direct the reader to \citet{Cluver2010} for a description of the observational details and data reduction. 
The upper right panel of Fig.~\ref{Fig_NUV_16_24_70mic_contours} shows the $16\,\mu$m data from the \textit{IRS} blue \textit{PUI}. The bottom left and right panels show the \textit{MIPS} $24$ and $70\,\mu$m images.

The SQ shock region was also mapped with the \textit{IRS} spectrograph \citep{Cluver2010}. The Short-Low (SL) and Long-Low (LL) modules of the spectrograph have been used, covering the wavelength ranges $5.3-14.0$ and $14-38$~$\mu$m, with spectral resolution of $\lambda / \delta \lambda = 60-127$ and $57-126$, respectively. 

\subsection{Ancillary data}
\label{ancillary-data}

We present the UV, optical and near-IR ancillary data  we use to determine 
the radiation ﬁeld heating the dust. (sect.~\ref{subsec:excitingfields}). 

\subsubsection{\textit{GALEX} UV imaging}
\label{UVimaging}

  \begin{figure*}
     \includegraphics[width = \textwidth]{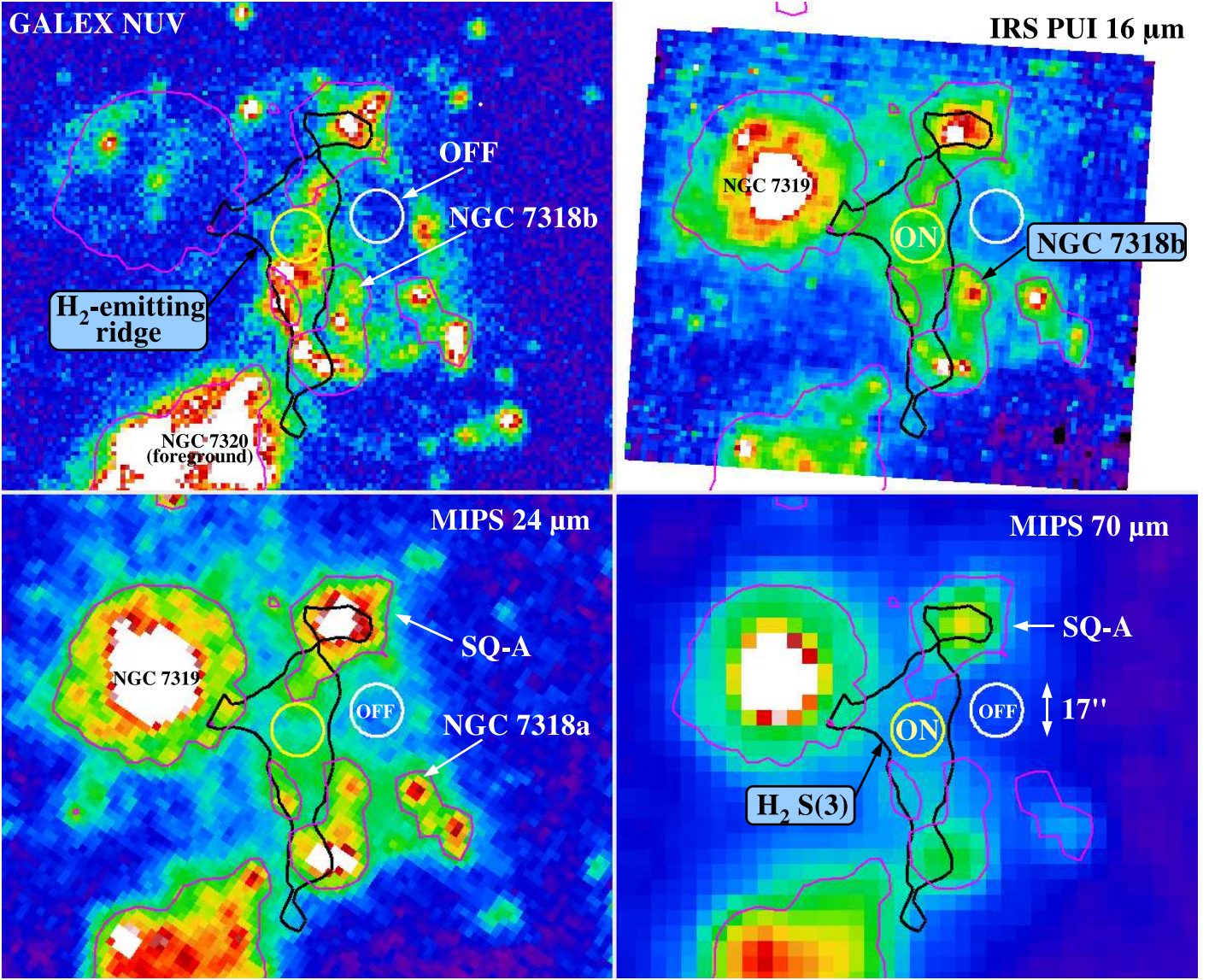}
     \caption{UV and mid-IR observations of Stephan's Quintet. \textit{Top-left} is a near-UV ($2\,267$~\AA) \textit{GALEX} image from \citet{Xu2005}, \textit{top-right} is the \textit{Spitzer IRS} $16\,\mu$m peak-up image, \textit{bottom-left} and \textit{bottom-right} are the $24\,\mu$m and $70\,\mu$m \textit{Spitzer MIPS} images of the SQ group, respectively. For all images, the \textit{black} contours shows the $4 \, \sigma$ (0.3~MJy~sr$^{-1}$) 0-0 S(3) H$_2$ line emission detected over the SQ ridge, from \textit{Spitzer IRS} mapping by \citet{Cluver2010}. The \textit{magenta}  contours show the $24\,\mu$m emission at a $0.25$~MJy~sr$^{-1}$ level. They are used to identify star forming regions that overlap the H$_2$ contours in the ridge. Photometry is performed within the H$_2$ contours, excluding or not these star-forming regions. The circles indicate 17" beams  where aperture photometry is also performed. The yellow circle shows the ``ON'' position in the SQ shock, centered on $\alpha = 22{\rm h}35'59.8''$, $\delta = +33\degr 58'16.7''$. The white circle show the  ``OFF'' position we use to subtract the background signal, centered on  $\alpha = 22{\rm h}35'57.7''$, $\delta = +33\degr 58'23''$.
}
  \label{Fig_NUV_16_24_70mic_contours}
  \end{figure*}

The upper left corner of Fig.~\ref{Fig_NUV_16_24_70mic_contours} shows the near-UV (NUV) image from the \textit{Galaxy Evolution Explorer} mission \citep[\textit{GALEX,}][]{Martin2005}. These observations have been reported by \citet{Xu2005}.
The pixel size is $1.5"$ , the wavelength is $\lambda = 2\,267$~\AA~and the bandwith is $\Delta \lambda = 732$~\AA. The far-UV (FUV) image ($\lambda = 1\,516$~\AA, $\Delta \lambda = 268$~\AA) is also used but is not shown here.

If one excludes the foreground galaxy \object{NGC 7320} (Sd), most of the UV emission is associated with the two spiral members of the group, NGC~7319 and NGC~7318b, and the intragroup medium starburst SQ-A. The galaxy-wide shock structure, which shows up in H$\alpha$, radio and X-ray observations, is barely visible on the UV images. 
Based on the comparison between \textit{ISO} and \textit{GALEX} data, \citet{Xu2005} concluded that most of the UV emission in the ridge is not associated with the large-scale shock itself, but with H$\,${\sc ii} regions along the spiral arm of the intruder NGC~7318b. 


We use the flux calibration described in the \textit{GALEX} observer's guide\footnote{\url{http://galexgi.gsfc.nasa.gov/docs/galex/}} and in \citet{Morrissey2007}. The unit data number (DN, or 1 count per second, cps) is equivalent to $108$ and $36\,\mu$Jy for FUV and NUV, or, equivalently, $1.4 \times 10^{-15}$ and $2.06 \times 10^{-16}$~erg~s$^{-1}$~cm$^{-2}$~\AA$^{-1}$, respectively.
We do not apply any aperture correction for the extended source UV photometry on \textit{GALEX} images since the PSF full width half maxima (FWHM) are 4.9'' and 4.2'' for the NUV and FUV, respectively.

\subsubsection{HST V-band imaging with WFPC2}
\label{subsec:HST_imaging}

We use $V$-band data taken with the \textit{Wide Field Planetary Camera 2 (WFPC2)} onboard the \textit{Hubble Space Telescope (HST)}.  For the F569W filter, the data consist of two sets of two images ($4 \times 800$~s exposure in total), each set for a given dithering position. The data were first processed by the \textit{HST} pipeline. After alignment of the images with the IRAF data reduction software, a median combination of the four images has been taken, followed by the removal of the remaining hot pixels by a $3 \times 3$ pixels median filter. We obtained a $V$-band  image similar to that presented in \citet{Gallagher2001}.

The photometry has been performed using the flux calibration given in the header of the images (1 DN$=4.149 \times 10^{-21}$~W~m$^{-2}$~\AA$^{-1}$). The central wavelength of the F569W $V$-band filter is  $\lambda = 5\,644.4$~\AA.

\subsubsection{Near-IR \textit{WIRC} imaging}
\label{subsec:nearIRimaging}


We use near-IR (NIR) data from recent deep observations with the Wide field IR Camera (\textit{WIRC}) on the Palomar 200-inch telescope (V. Charmandaris, private communication). 
The \textit{WIRC} images were taken in July, 2009, and processed with the \textit{Swarp} software\footnote{\url{http://astromatic.iap.fr/software/swarp}}. They are $5-6$ magnitudes deeper than the corresponding \textit{2MASS} images. The zero-point magnitudes are $24.50$, $22.73$ and $23.05$, so we use flux calibrations of $0.252$, $0.830$, and $0.402$~$\mu$Jy~DN$^{-1}$ for J, H, and K$_{\rm s}$, respectively. 
The corresponding central wavelengths used are $1.235 \pm 0.006$, $1.662 \pm 0.009$, $2.159 \pm 0.011$~$\mu$m. 
The images show that most of the NIR emission in the SQ ridge is associated with the spiral arm of the intruder, NGC~7318b. NIR emission associated with the SQ-A northern starburst is also detected.




\section{Photometry on IR, optical and UV images}
\label{sec:photometry}

We describe the method used to perform the photometry on \textit{Spitzer} mid-IR images for the dust emission and on near-IR, optical and UV images to estimate the radiation field at the position of the SQ shock. 

\subsection{Method}

\subsubsection{Dust emission}
\label{photometry_method_dust_emission}

The SQ field of view is a crowded region (see sect.~\ref{spatial-distribution-dust-emission} for a description of the spatial distribution of the dust emission). 
In order to isolate the dust emission from the shock itself, we sum the signal within regions that are not contaminated by IR-emitting, star-forming regions.  
Since the shock is surrounded by bright sources, we do not apply any aperture correction.
For comparison, three different area are used to perform photometric measurements.
\begin{enumerate}
\item The signal is summed over a circular aperture of $17"$ in diameter that is centered on the SQ ridge, in the middle of the X-ray emitting shock front. This $17"$ aperture corresponds to the FWHM of the MIPS beam at $\lambda = 70\,\mu$m. It is marked with the yellow circle on Fig.~\ref{Fig_NUV_16_24_70mic_contours} (``ON'' position). 
\item The signal is integrated over the SQ ridge within the H$_2$ contours (black line on Fig.~\ref{Fig_NUV_16_24_70mic_contours}) that define the shock structure. 
Except SQ-A, star-forming regions within the black H$_2$ emission contour are included. They are defined by the $24\,\mu$m iso-flux ($F_{24\,\mu \rm m} > 0.25$~MJy~sr$^{-1}$) contour (magenta line in Fig.~\ref{Fig_NUV_16_24_70mic_contours}).
\item  The signal is summed over the SQ ridge within the H$_2$ contour but excluding the star forming regions. To do this, we subtract from the signal the emission arising from the intersection of the areas within the magenta $24\,\mu$m contours and black H$_2$ contours. 
\end{enumerate}

To remove the contamination from 
the halo of the group, we choose to estimate the background level in a region close to the H$_2$-emitting shock. This ``OFF'' region is marked with the white circle on Fig.~\ref{Fig_NUV_16_24_70mic_contours}.  For the $70\,\mu$m image, we cannot exclude that the flux within the ``ON'' position is contaminated  by the brighter  sources around it. Thus, we consider our $70\,\mu$m flux as an upper limit. A more detailed analysis of the \textit{MIPS} 70 and 160$\,\mu$m will be reported in \citet{Natale2010}.

\subsubsection{Radiation field}
\label{photometry_method_radiation-field}

The radiation field is the integral of the flux over all directions. In the UV domain, due to scattering of light, we can assume that the radiation  field is isotropic and estimate its strength from UV photometry at the position of the shock. Thus we measure the UV fluxes within the same apertures than for the \textit{Spitzer} images, using the same ``OFF'' position. 

The UV fluxes are corrected for both foreground galactic and SQ internal extinctions.
For the extinction curve of the Galactic diffuse interstellar medium \citep[$R_{\rm V}=3.1$ curve in][]{Weingartner2001}, the visible extinction $A_V^{\rm (MW)} = 0.24$ scales to the FUV and NUV Galactic extinctions at \textit{GALEX} wavelengths, respectively $A_{\rm FUV}^{\rm (MW)} = 0.62$ and $A_{\rm NUV}^{\rm (MW)} = 0.72$. We use the \citet{Xu2005} values for the internal SQ extinction, i.e.  $A_{\rm FUV}^{\rm (SQ)} = 0.76$ and $A_{\rm NUV}^{\rm (SQ)} = 0.88$.

The optical and near-IR images show that the shock is surrounded by bright sources, in particular NGC~7318b and nearby star-forming regions. Since there is little scattering at these wavelengths, the photometry restricted to the shock area is likely to underestimate the optical and near IR intensity of the radiation field. The choice of the aperture is rather arbitrary in that case and only provide a rough approximation of the optical and near-IR radiation field. Note that the photometry has been performed after removal of Galactic stars (using the DAOPHOT package).  

Based on an average optical extinction of $A_V = 0.6$ \citep{Guillard2009} for the center position in the ridge, and a Galactic extinction curve, we apply the following extinction corrections: $A_{\rm J} = 0.17$, $A_{\rm H} = 0.11$, and $A_{\rm K_s} = 0.07$ for J, H, and K$_{\rm s}$ bands, respectively. We use the values of zero-point fluxes from \citet{Cohen2003}, i.e. $1594 \pm 28$, $1024 \pm 20$, $666.7 \pm 12.6$~Jy for J, H, and K$_{\rm s}$ bands. The internal extinction correction applied on the \textit{HST} photometry is $A_{F569W} = 0.575$.




\renewcommand{\arraystretch}{1.2} 
\begin{table*}
\begin{center}
\begin{minipage}[t]{\textwidth}
 \renewcommand{\footnoterule}{}
\def\thefootnote{\alph{footnote}}
\centering
 \caption[]{Summary of the mid-IR photometric measurements\footnotemark[1]  performed on the \textit{Spitzer IRAC}, \textit{IRS PUI} $16\,\mu$m, and \textit{MIPS} $24$, and $70\, \mu$m images (Fig.~\ref{Fig_NUV_16_24_70mic_contours}). The background-subtracted surface brightnesses are indicated for three different apertures. The last row ($17"$ circular aperture ) shows the values we adopt throughout this paper, and in particular for the SED presented in Fig.~\ref{Fig_SED_Galactic_SQshock}.}
    \begin{tabular}{c c c c c c c c}
	\hline
	\hline
   Target region & $3.6\,\mu$m & $4.5\,\mu$m & $5.8\,\mu$m & $8\,\mu$m  \footnotemark[5] &  $16\, \mu$m \footnotemark[6] &  $24\, \mu$m &$70\, \mu$m \\
\hline
 SQ ridge \footnotemark[2]                 & $ 23.7\pm0.9$ & $ 11.9\pm0.2$ & $ 7.3\pm0.2$ & $4.2\pm0.1$ & $1.25\pm 0.05$ & $2.25\pm0.12$ & $<6.43$  \\                                                          
 \hline
SQ ridge (partial) \footnotemark[3]     & $ 19.4\pm1.7$  & $9.9\pm1.8$  & $ 6.0\pm0.1$ &$ 3.35\pm0.11$ & $1.00\pm 0.06$ & $0.89\pm0.15$ &  $<5.04$  \\
 \hline
$17"$ beam \footnotemark[4]            &$ 17.6\pm0.9$ & $9.15\pm0.12$ & $ 5.29\pm 0.15$ & $3.07\pm0.11$  &  $0.88\pm0.03$ & $0.79\pm0.15$ & $<3.91$ \\
 \hline
\end{tabular}
\label{table_midIR_fluxes}
\footnotetext[1]{Fluxes are in units of $\times 10^{-8}$~W~m$^{-2}$~sr$^{-1}$.}
\footnotetext[2]{Mid-IR signal integrated over the shock structure, defined by the 4$\,\sigma$ S(3) H$_2$ rotational line contours (black line on Fig.~\ref{Fig_NUV_16_24_70mic_contours}). We exclude the SQ-A star-forming region.}
\footnotetext[3]{Mid-IR signal integrated over the S(3) H$_2$ rotational line contours by excluding overlaps with the ($\mathcal{F}_{24\,\mu\rm m} > 0.25$~MJy~sr$^{-1}$) $24\,\mu$m magenta contours (see  Fig.~\ref{Fig_NUV_16_24_70mic_contours}). This allows us to partially remove the contribution from star-forming regions to the dust emission.}
\footnotetext[4]{Mid-IR signal integrated over the 17'' beam in the center of the SQ shock structure (yellow circle on Fig.~\ref{Fig_NUV_16_24_70mic_contours}) }
\footnotetext[5]{The contribution of the  $8\,\mu$m S(4) H$_2$ line emission to the $8\, \mu$m in-band flux is subtracted.}
\footnotetext[6]{The contribution of the  $17\,\mu$m S(1) H$_2$ line emission to the $16\, \mu$m in-band flux is subtracted.}
\end{minipage}
\end{center}
\end{table*}
\renewcommand{\arraystretch}{1} 

\subsection{Results}
\label{photometry:results}

The quantitative results about the photometry performed on IR \textit{Spitzer} images are gathered in Table~\ref{table_midIR_fluxes}. 
The surface brightnesses are given for the three area described above. 
We estimate from the \textit{IRS} spectrum that the $17\,\mu$m S(1) and $8\,\mu$m S(4) H$_2$ line emission represent respectively $68\,$\% and $62\,$\% of the \textit{IRS} Peak-Up Imager $16\,\mu$m and the \textit{IRAC} $8\,\mu$m in-band flux within the $17''$ ON aperture centered on the shock, and we correct for this contamination.
The last row of table~\ref{table_midIR_fluxes} indicates the fluxes we adopt throughout this paper. The error bars are estimated by using two different background subtractions close to the ``OFF'' position (8-arcsec shifts in the East-West direction). In addition, we include calibration uncertainties (of the order of 5\%).


\renewcommand{\arraystretch}{1.2} 
\begin{table*}
\begin{center}
\begin{minipage}[t]{\textwidth}
 \renewcommand{\footnoterule}{}
\def\thefootnote{\alph{footnote}}
\centering
  \caption[]{Summary of the UV photometric measurements performed on \textit{GALEX} FUV and NUV images (see Fig.~\ref{Fig_NUV_16_24_70mic_contours}). The flux columns indicate surface brightnesses that are corrected for the sky background.}
    \begin{tabular}{c|c|c|c|c}
	\hline
	\hline
      \multirow{3}*{Target position} &  \multicolumn{2}{|c|}{FUV ($\lambda = 1\,516$~\AA)} &   \multicolumn{2}{|c}{NUV ($\lambda = 2\,267$~\AA)} \\
\cline{2-5}
	& Flux & Radiation Field $G$ &  Flux & Radiation Field $G$ \\
	& $\times 10^{-7}$ [ W m$^{-2}$ sr$^{-1}$]  & [Habing units] &  $\times 10^{-7}$ [W m$^{-2}$ sr$^{-1}$] & [Habing units] \\
\hline
 SQ ridge \footnotemark[1]                & $2.82 \pm 0.34$ & $1.54\pm0.19$ & $2.74\pm0.39$ & $1.98\pm0.29$  \\
	\hline
 SQ ridge (partial) \footnotemark[2]     & $2.29 \pm 0.35$ & $1.25 \pm 0.19$ & $2.25 \pm 0.37$ & $1.63 \pm 0.29$ \\
	\hline
$17"$ ON beam \footnotemark[3]            & $2.49 \pm 0.35$ & $1.36 \pm 0.19$ & $2.42 \pm 0.39$ & $1.75 \pm 0.29$  \\
  \hline
	\hline
    \end{tabular}
 
    \label{table_UVfluxes}
\footnotetext[1]{UV signal integrated over the shock structure, defined by the 4$\,\sigma$ S(3) H$_2$ rotational line contours (black line on Fig.~\ref{Fig_NUV_16_24_70mic_contours}). We exclude the SQ-A star-forming region.}
\footnotetext[2]{UV signal integrated over the S(3) H$_2$ rotational line contours by excluding overlaps with the ($\mathcal{F}_{24\,\mu\rm m} > 0.25$~MJy~sr$^{-1}$) $24\,\mu$m magenta contours (see  Fig.~\ref{Fig_NUV_16_24_70mic_contours}). This allows us to partially remove the contribution from star-forming regions to the dust emission.}
\footnotetext[3]{UV signal integrated over the 17'' beam in the center of the SQ ridge (yellow circle on Fig.~\ref{Fig_NUV_16_24_70mic_contours}) }
\end{minipage}
\end{center}
\end{table*}
\renewcommand{\arraystretch}{1}

\renewcommand{\arraystretch}{1.1} 
\begin{table*}
\begin{center}
\begin{minipage}[t]{\textwidth}
 \renewcommand{\footnoterule}{}
\def\thefootnote{\alph{footnote}}
\centering
 \caption{Optical \textit{HST} $V$-band and near-infrared \textit{WIRC} surface brightnesses\protect\footnotemark[1] in the J, H and K$_{\rm s}$ bands for the SQ ridge.}
    \begin{tabular}{c c c c c c c}
	\hline
	\hline
      \multicolumn{2}{c}{Band} & V & J & H & K$_{\rm s}$ & J:H:K$_{\rm s}$ \\
	\hline
  & Flux   & $7.2\pm 1.6$ & $ 8.45 \pm 0.64$ & $ 8.54 \pm 0.65$ & $8.07 \pm 0.61$ &   1.048: 1.058 :  1 \\
    & Dered & $12.2 \pm 2.9$ & $9.93 \pm 0.77$ & $9.44 \pm 0.77 $ & $8.60 \pm 0.75$ & 1.154 : 1.096 : 1 \\
\hline
	\hline
\end{tabular}
\label{table_2MASSfluxes}
\footnotetext[1]{The fluxes are given in units of $10^{-7}$~W~m$^{-2}$~sr$^{-1}$. ``Dered'' indicates the extinction-corrected brightness, and the ``J:H:K$_{\rm s}$'' column shows flux ratios between the bands.}
\end{minipage}
\end{center}
\end{table*}
\renewcommand{\arraystretch}{1}

The results of the UV \textit{GALEX} photometry are gathered in Table~\ref{table_UVfluxes}.
The surface brightnesses are given for the three apertures defined in sect.~\ref{photometry_method_dust_emission}. 
The UV fluxes measured over the SQ ridge aperture that includes the $24\,\mu$m-bright ($> 0.25$~MJy~sr$^{-1}$) regions are $10-25$~\% higher than the fluxes where these regions have been removed from the aperture (``SQ ridge (partial)'' row). 
The brightnesses are background-subtracted and corrected for foreground galactic extinction and SQ internal extinction.
From the UV surface brightness values we derive the intensity  of the standard interstellar radiation field in  Habing units.


The results show that the UV flux in the shock, outside star-forming regions,  corresponds to an interstellar radiation field of average intensity $G_{\rm UV}=1.4 \pm 0.2$ in Habing units\footnote{The flux of the Habing field ($G_{\rm UV}=1$) equals $2.3 \times 10^{-3}$~erg~s$^{-1}$~cm$^{-2}$ at $\lambda =  1\,530$~\AA ~\citep{Habing1968,Mathis1983}.} (see sect.~\ref{subsubsec:ISRF} for details). The value of $G_{\rm UV}$ is used to characterize the intensity of the non-ionizing radiation field in the shock (sect.~\ref{subsec:excitingfields}). The error bar on the $G_{\rm UV}$ factor takes into account two background subtractions obtained by shifting the OFF position by 8'' in the East-West direction. 
This uncertainty does not take into account the uncertainty on the optical extinction. Although optical spectroscopy shows that there are spatial variations of the $A_{\rm V}$ value in the shock ($A_{\rm V}=0.1-2.5$), we do not expect that this has an important impact on our estimate of $G_{\rm UV}$, because it is calculated over a large aperture. If the most extincted regions have a reasonably small surface filling factor, they will not affect $G_{\rm UV}$ significantly.


The extinction-corrected surface brightnesses measured on \textit{HST} $V$-band and \textit{Palomar WIRC} images are gathered in Table~\ref{table_2MASSfluxes}.

\section{Dust emission from Stephan's Quintet: observational results}
\label{sec:dustemission}


This section reports the observational results about the dust emission detected in the SQ ridge.
 We compare the mid-IR SED and the relative intensities of the PAH bands in the SQ shock to that of the Galactic diffuse interstellar medium (ISM).

\subsection{Spatial distribution of the dust emission from the SQ group}
\label{spatial-distribution-dust-emission}

The images in Fig.~\ref{Fig_NUV_16_24_70mic_contours} show that the bright mid-IR 
emitting regions are associated with the NGC 7319 galaxy and with UV-luminous, star-forming regions \citep{Cluver2010}. The mid-IR emission does not correlate with the H$_2$, X-ray  and  radio emissions. Interestingly, these star-forming regions are outside the galactic disks of NGC~7318a and b, suggesting that a significant amount of gas has been displaced from these galaxies by tidal interactions.
If we exclude the foreground galaxy NGC~7320, the bright mid-IR emission comes from the NGC 7319 galaxy, the SQ-A starburst region, and an arc-like feature to the east of the intruding galaxy NGC~7318b. This arc structure  is clearly seen on the UV, \textit{IRAC}, and $16\,\mu$m images, and could be associated with NGC~7318b's spiral arm. The $24$ and $70\,\mu$m images show a bright and extended emission at the southern tip of the ridge which may be associated with enhanced star formation in this arc feature. 

Dust is detected in the SQ shock region, at a distance of $10-20$~kpc of the nearest surrounding galaxies. 
In the following, we focus on this faint emission observed within the SQ ridge, outside star-forming regions. 
Note that on the 16$\,\mu$m \textit{IRS Blue Peak-Up}  image, the emission in the ridge correlates well with the H$_2$ emission, which is due to the contamination of the H$_2$ 17$\,\mu$m S(1) line to the in-band flux (sect.~\ref{photometry:results}).

Using a combination of the H$\alpha$ and 24$\,\mu$m luminosities, or the 7.7$\,\mu$m PAH emission, \citet{Cluver2010} find an upper limit of $\lesssim 0.08$~M$_{\odot}$~yr$^{-1}$ on the star formation rate in the shock, as compared with $\sim 1.25$~M$_{\odot}$~yr$^{-1}$ in SQ-A. 
This  suggests that the star formation is being depressed in the shock region. 

\subsection{Dust emission from the SQ shock, outside star-forming regions}
\label{dust-emission-from-SQ-shock}

\begin{figure}
     \centering
     \includegraphics[angle=90, width = \columnwidth]{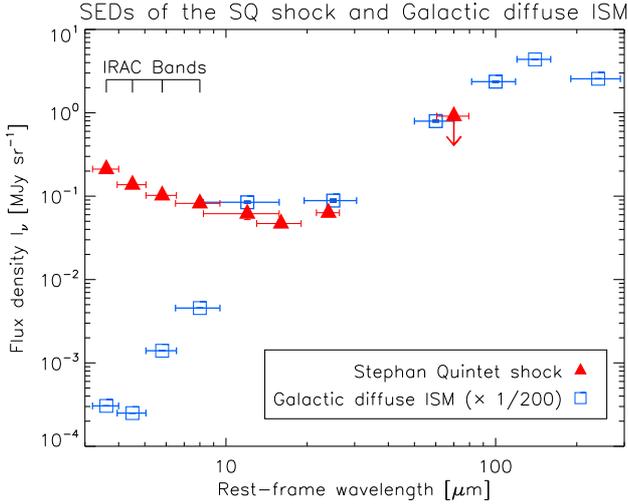}
 \caption{The SED of the dust emission from the SQ shock seen by \textit{Spitzer} (table~\ref{table_midIR_fluxes}), compared with that of the Galactic diffuse emission observed with \textit{Spitzer IRAC}, \textit{IRAS} and \textit{DIRBE} towards the line-of-sight centered on the Galactic coordinates $(28.6, +0.8)$. The contribution of the gas lines is subtracted. The Galactic SED is scaled down by a factor $200$. Between $3-8\,\mu$m, the SQ flux is much brighter because of the stellar contribution. Horizontal bars indicate the filters bandwiths.}
     \label{Fig_SED_Galactic_SQshock}
  \end{figure}

The images and photometric measurements show that thermal dust emission is detected in the SQ shock structure, outside star-forming regions. The SED of the IR emission from the center of the shock is shown on Fig.~\ref{Fig_SED_Galactic_SQshock}. Among the SEDs listed in table~\ref{table_midIR_fluxes}, the figure displays that of the last row (17'' beam, see sect.~\ref{sec:photometry}). 

For comparison with Galactic data, the figure includes a 12$\,\mu$m flux  computed by integrating the \textit{IRS} spectrum over the IRAS 12$\,\mu$m filter bandpass after  subtraction of the gas lines. 
The $12 / 24\,\mu$m flux density ratio in the SQ shock is $\mathcal{F}_{\nu}^{\rm (SQ)}(12\,\mu m) /   \mathcal{F}_{\nu}^{\rm (SQ)}(24\,\mu m) = 0.97 \pm 0.15$, which is remarkably close to the value for the Galactic diffuse ISM: $\mathcal{F}_{\nu}^{\rm (MW)}(12\,\mu m) /   \mathcal{F}_{\nu}^{\rm (MW)}(25\,\mu m) = 0.95 \pm 0.07$. 
The Galactic SED is that measured towards the line-of-sight centered on the Galactic coordinates $(28.6, +0.8)$, observed with the ISOCAM-CVF\footnote{Camera equiped with a Circular Variable Filter (CVF) on-board the \textit{Infrared Space Observatory} (ISO)} and \textit{Spitzer IRAC} \citep{Flagey2006}. Fig.~\ref{Fig_SED_Galactic_SQshock} also shows the \textit{IRAS} 12, 25, 60 and 100$\,\mu$m fluxes for the Galactic diffuse ISM emission measured on \textit{IRIS} images \citep{Miville-Deschenes2005a}. The  \textit{IRAS} SED is extended to 140 and 240$\,\mu$m using  \textit{DIRBE} \citep{Hauser1998} color ratios at the position of the line of sight. The total column density for this line of sight is estimated to be $N_{\rm H} = 2 \times 10^{22}$~cm$^{-2}$, and the mean radiation field a few Habing units ($G_{\rm UV} \sim 3$). In the figure, the Galactic SED is scaled down by a factor $200$. 



We propose that the faint dust emission we have isolated in the center of the SQ shock is diffuse emission associated with the shocked molecular gas present in the ridge. An second possibility is that the dust is associated with the H{\sc i} gas  and 
a third one that it is produced by unresolved star-forming regions.
We favor the first interpretation for three reasons.

\begin{itemize}

\item

Looking at the H{\sc i} data of \citet{Williams2002}, the  two outer contours in their figure 9, $0.6$ and $1\times 10^{20}$~cm$^{-2}$, intersect our 17'' aperture used for dust photometry. Since the angular resolution of these H{\sc i} observations is 20'', the H{\sc i} emission is likely to be  contaminated by beam smearing of  the brighter emission to the north of our aperture. On the southern side of our aperture, in the ridge, H{\sc i} is undetected and $N_{\rm HI} < 5.8 \times 10^{19}$~cm$^{-2}$. This upper limit is smaller than the  column density of warm H$_2$ ($2\times 10^{20}$~cm$^{-2}$) derived from the \textit{Spitzer} H$_2$ fluxes. Since the warm H$_2$ column density is a lower limit on the total H$_2$ column density, 
H{\sc i} gas accounts for a minor fraction of the gas column density  in the ridge. Therefore the dust emission cannot be mostly associated with the H{\sc i} gas in the shock region.

\item

\textit{HST} observations show that there are very few star clusters in the center of the ridge. Most of them are associated with NGC~7319, the tidal debris of NGC 7318a/b, and the SQ-A intragroup starburst region \citep{Gallagher2001}. We find three candidates for star clusters ($M_{\rm V} = -12.14, \ -10.01, \ -9.65$) within our $17''$ ($\sim 7.8$~kpc) aperture in the center of the shock. These three clusters produce a $V$-band flux of $5.2 \times 10^{-9}$~W~m$^{-2}$~sr$^{-1}$, which is 2 orders of magnitude lower than the $V$-band surface brightness that we derived from $HST$ observations in the center of the ridge. 

\item The $12 / 24\,\mu$m flux ratio is remarkably similar to that of the Galactic diffuse ISM.
The average column density of warm molecular gas in the SQ shock derived from H$_2$ observations is $N_{\rm H} = 2 \times 10^{20}$~cm$^{-2}$, a factor 100 smaller than that of the Galactic line of sight, in agreement with the scaling factor used to match the fluxes in Fig.~\ref{Fig_SED_Galactic_SQshock}.

\end{itemize}

\subsection{Mid-IR spectroscopy: characterization of the dust emission from the SQ shock}
\label{subsec:IRS}

\subsubsection{\textit{Spitzer IRS} spectrum in the center of the shock}
\label{subsubsec:IRSdata}

  \begin{figure*}
    \centering
     \includegraphics[height = 0.75\textwidth, angle=90]{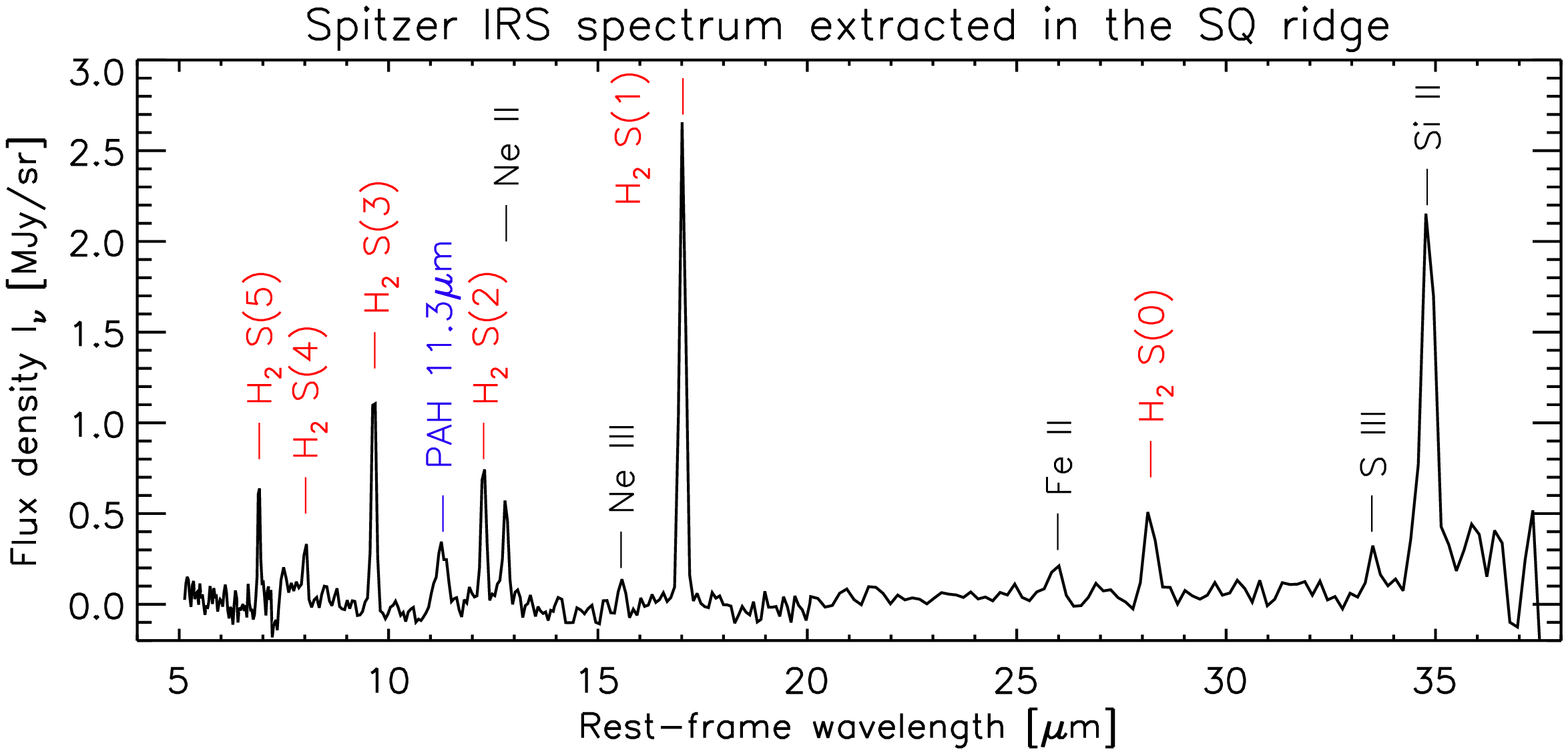}
     \includegraphics[height = 0.75\textwidth, angle=90]{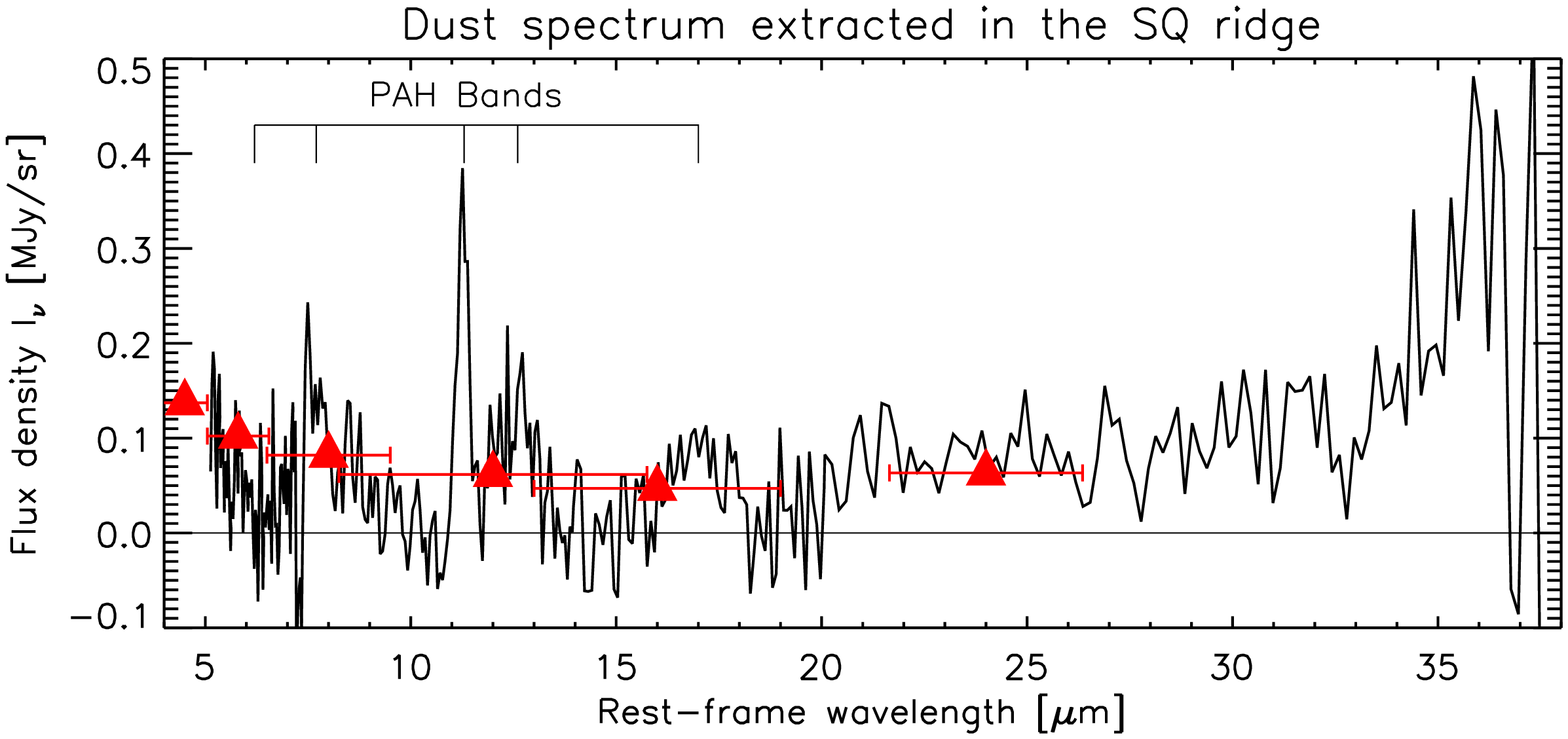}
 \caption{\textit{Spitzer IRS} mid-IR spectra extracted over a $\sim 18" \times 15"$ area in the center of the SQ shock structure (RA 22:35:59.9, DEC +33:58:16.8). This aperture corresponds approximately to the ``ON'' beam shown in Fig.~\ref{Fig_NUV_16_24_70mic_contours}. \textit{Top:} full spectrum. \textit{Bottom:} gas lines are removed to highlight PAH dust features and thermal dust continuum. Photometric measurements (red triangles) performed on mid-IR images (same as in Table~\ref{table_midIR_fluxes} and Fig.~\ref{Fig_SED_Galactic_SQshock}) are overplotted. Horizontal bars indicate the filters bandwiths.
}
     \label{Fig_IRS_spec_SQ_shock}
  \end{figure*}

  \begin{figure*}
     \centering
     \includegraphics[width = 0.75\textwidth]{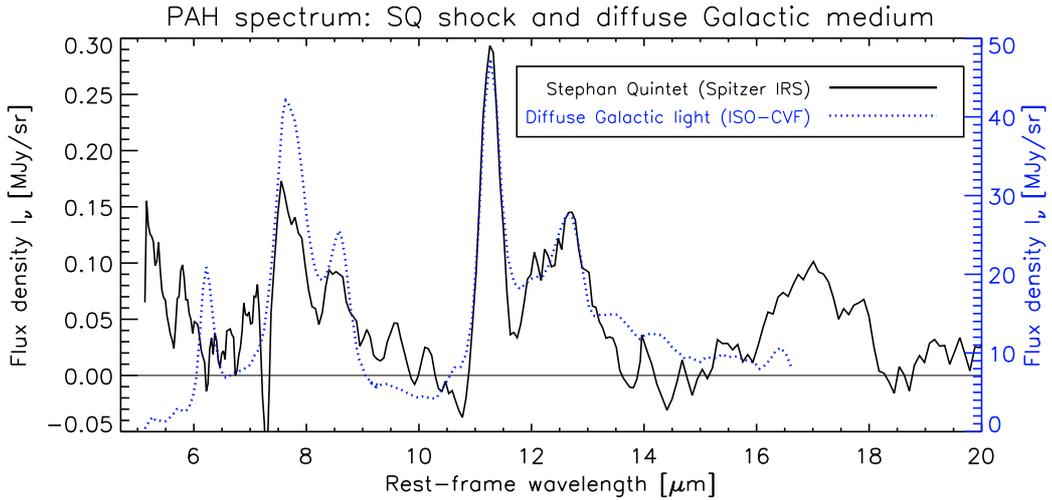}
 \caption{\textit{Spitzer IRS} PAH emission spectrum extracted in the center of the SQ ridge (black solid line), compared with  the \textit{ISOCAM-CVF} spectrum of the diffuse galactic medium (blue dashed line, with flux density labeled on the right). The \textit{IRS} spectrum is smoothed to a resolution $\mathcal{R} = \lambda / \delta \lambda = 24 \rightarrow 51$, comparable with the \textit{ISO-CVF} resolution ($\mathcal{R} = 35 \rightarrow 45$). Gas lines are removed in both spectra. Note that the flux scales are different for each spectrum (labels on the left for SQ, and on the right for the Galaxy). }
     \label{Fig_SQ_Galactic_PAH_comp}
  \end{figure*}

The top panel of Fig.~\ref{Fig_IRS_spec_SQ_shock} presents a low-resolution \textit{Spitzer IRS} spectrum extracted from a central position of the SQ ridge. 
This spectrum was obtained by \citet{Cluver2010}, averaging all the spectra observed within a $274.2\ \rm arcsec^{2}$ ($\sim 18" \times 15"$) rectangular aperture around the center of the ON position. These new data has much higher sensitivity and better flux calibration than the first observations reported by \citet{Appleton2006}.
This spectrum confirms the first results discussed in \citet{Appleton2006}. It shows bright S(0) to S(5) H$_2$ rotational lines  and forbidden atomic lines, with a remarkable $34.8\,\mu$m [Si$\,${\sc ii}] feature. Here we focus on the dust emission.

The \textit{IRS} spectrum shows that PAH and thermal dust emission is detected from the center of the SQ ridge, outside star-forming regions. 
 A weak thermal continuum is also visible from $20$ to $38\,\mu$m. The ratio between the fluxes of the $17\,\mu$m S(1) H$_2$ line and the $7.7\,\mu$m PAH feature is $1.2 \pm 0.2$ in the shock, which is about two orders of magnitude higher than the value observed in star-forming galaxies \citep{Roussel2007}.

\subsubsection{PAH emission from the SQ shock}
\label{subsubsec:PAH}

The $7.7$, $11.3$ and $17\,\mu$m Aromatic Infrared Bands (AIBs), attributed to PAHs, are detected (although weak)  in the new  \citet{Cluver2010} observations at the center of the SQ shock.
We use the PAHFit IDL tool\footnote{available on \url{http://tir.astro.utoledo.edu/jdsmith/pahfit.php}}  \citep{Smith2007} for decomposing the full $5-38\,\mu$m  \textit{IRS} spectrum  into contributions from PAH features, thermal dust continuum, starlight and gas lines. We do not include any extinction in the fit. The result of the fit on the full $5-38\,\mu$m spectrum is shown in Appendix~\ref{appendix_PAHFit}, Fig.~\ref{Fig_pahfit_noext_full_SQ}. This spectral decomposition allows us to remove the gas lines and extract a ``pure'' dust spectrum, shown on the bottom panel of Fig.~\ref{Fig_IRS_spec_SQ_shock}. PAHFit is run one more time on this spectrum, which allows us to measure accurately the fluxes of the AIBs. The line strengths of the PAH emission features and their ratios are gathered in Tables~\ref{table_PAHfluxes} and \ref{table_PAHratios}, respectively.

In Fig.~\ref{Fig_SQ_Galactic_PAH_comp} we compare the PAH spectrum from the SQ shock with the \textit{ISOCAM-CVF} spectrum of the diffuse Galactic emission from \citet{Flagey2006}, between $5$ and $20\,\mu$m. 
The main differences between the two spectra are the following. First, the ratio between the flux in the band at $7.7\,\mu$m and the band at $11.3\,\mu$m (henceforth $R_{7.7/11.3}$)  is a factor $\sim 2$ lower for SQ ($R_{7.7/11.3}^{(SQ)} = 1.36 \pm 0.07$, see table~\ref{table_PAHratios}) than for the diffuse Galactic light ($R_{7.7/11.3}^{(MW)} = 2.93 \pm 0.08$). 
The $R_{7.7/11.3}$ value for the SQ shock is comparable to that measured for AGN of the \textit{SINGS}\footnote{Spitzer Infrared Nearby Galaxies Survey, \url{http://sings.stsci.edu/}.} sample \citep{Smith2007}. 
Second, the $6.2\,\mu$m AIB is absent in the SQ spectrum. Note that the rise of the SQ spectrum at $\lambda < 6\,\mu$m is due to the stellar component.
The $17\,\mu$m complex is prominent but the $16.4\,\mu$m feature is not seen on top of it. This may not be significant because, to our knowledge, this feature is only prominent in  the NGC~7023 Galactic PDR\footnote{Photo-Dissociation Region} \citep{Sellgren2007}.  This feature is also observed in the star-forming galaxy NGC~7331 \citep{Smith2004}, but with a higher signal to noise spectrum than the one we have for SQ.

\renewcommand{\arraystretch}{1.2} 
\begin{table*}
\begin{center}
\begin{minipage}[t]{\textwidth}
 \renewcommand{\footnoterule}{}
\def\thefootnote{\alph{footnote}}
\centering
 \caption{Fluxes of the PAH bands measured with the PAHFit IDL tool on the \textit{Spitzer IRS} spectrum of the center of the SQ ridge and the  \textit{ISO-CVF} spectrum of the Galactic light.}
    \begin{tabular}{c c c c c c}
	\hline
	\hline
     &  \multicolumn{5}{c}{Integrated fluxes (see notes for units)}  \\
	\hline
     PAH band &  $6.2\,\mu$m  &  $7.7\,\mu$m \footnotemark[3] & $11.3\,\mu$m \footnotemark[3] & $12.6\,\mu$m \footnotemark[3] & $17\,\mu$m \footnotemark[3]  \\ 
	\hline
     SQ ridge \textit{(IRS)} \footnotemark[1] &  $< 0.75$ &  $3.8 \pm 0.1$ &  $2.8 \pm 0.1$ &  $1.2 \pm 0.2$ &  $0.93 \pm 0.04$ \\
	\hline
     Galactic \textit{(CVF)}  \footnotemark[2] & $5.19 \pm 0.09$ & $19.05 \pm 0.03$ & $6.49 \pm 0.17$ & $2.90 \pm 0.05$ & No data \\
	\hline
\end{tabular}
\label{table_PAHfluxes}
\footnotetext[1]{Fluxes are in units of $\times 10^{-9}$~[W~m$^{-2}$~sr$^{-1}$]}
\footnotetext[2]{Fluxes are in units of $\times 10^{-7}$~[W~m$^{-2}$~sr$^{-1}$]}
\footnotetext[3]{In the case of PAH blended complexes, we sum the fluxes of the different lorentzian components that contribute to the feature.}
\end{minipage}
\end{center}
\end{table*}
\renewcommand{\arraystretch}{1} 

\renewcommand{\arraystretch}{1.2} 
\begin{table*}
\begin{center}
\begin{minipage}[t]{\textwidth}
 \renewcommand{\footnoterule}{}
\def\thefootnote{\alph{footnote}}
\centering
 \caption{Flux ratios between the PAH bands for the \textit{Spitzer IRS} spectrum of the center of the SQ ridge and the  \textit{ISO-CVF} spectrum of the diffuse Galactic light.}
 \begin{tabular}{c c c c c}
	\hline
	\hline
 & \multicolumn{4}{c}{Flux ratios} \\
\hline
& $R_{6.2/7.7}$ & $R_{7.7/11.3}$  & $R_{11.3/12.6}$ & $R_{11.3/17}$ \\ 
\hline
 SQ ridge \textit{(IRS)} & $< 0.2 $ & $1.36 \pm 0.07$ & $2.3 \pm 0.2$ & $3.01 \pm 0.16$ \\ 
Galactic \textit{(CVF)} & $0.272 \pm 0.005$ & $2.93 \pm 0.08$ & $2.24 \pm 0.07$ &  No data \\
\hline
\end{tabular}
\label{table_PAHratios}
\end{minipage}
\end{center}
\end{table*}
\renewcommand{\arraystretch}{1}

The enhancement of the $11.3\,\mu$m AIB compared to the 6.2, 7.7 and 8.6$\,\mu$m features has been observed on galactic scales in elliptical galaxies  \citep[e.g.][]{Kaneda2005}, and in active galactic nuclei \citep[e.g.][]{Smith2007}. 
This enhancement has also been discussed  in the Galactic diffuse medium \citep[e.g.][]{Flagey2006} or at  small scales, in PDR interfaces \citep[e.g.][]{Rapacioli2005, Compiegne2007}. 

The PAH emission spectrum depends on the size distribution, hydrogenation and ionization states. The $R_{7.7/11.3}$ PAH band ratio depends mainly on the PAH ionization state. Theoretical \citep{Langhoff1996, Bakes2001a, Bakes2001b, Draine2001, Bauschlicher2002} and experimental \citep[e.g.][]{Szczepanski1993} studies show that neutral PAHs have lower $R_{7.7/11.3\mu\rm m}$ than charged ones (both anions and cations).  The charge state of PAHs is mainly determined by the ionization parameter $G_{\rm UV} \sqrt{T} / n_{\rm e}$ where $G_{\rm UV}$ is the integrated far ultraviolet
($6-13.6$~eV) radiation field expressed in units of the Habing radiation field, $T$ is the electron  temperature and $n_{\rm e}$ is the electronic
density. This parameter translates the balance between photoionisation and recombination rates of electrons \citep{Bakes1994,  Weingartner2001b}. 
\citet{Flagey2006} quantified $R_{7.7/11.3\mu\rm m}$ as a function of the ionization parameter $G_{\rm UV} \sqrt{T} / n_{\rm e}$  and the PAH mean size. We use their calculations to discuss the ionization state of the PAHs in the SQ shock. Our PAHFit decomposition yields  $R_{7.7/11.3\mu\rm m} = 1.36 \pm 0.07$, which translates into $G_{\rm UV} \sqrt{T} / n_{\rm e} \lesssim  30$~K$^{1/2}$~cm$^{3}$. Assuming a warm molecular gas temperature of $\sim 150$~K and $G_{\rm UV}=1$, we find that $n_{\rm e} \gtrsim 0.4$~cm$^{-3}$. This lower limit  is one order of magnitude higher than the electronic densities inferred from observations and modeling of the ionization of cold neutral gas in the Solar neighbourhood \citep{Weingartner2001a}. If this interpretation and diagnostic applies, a higher ionizing flux from cosmic-rays or X-rays would be required to maintain such a high electron density.

The $R_{6.2/7.7}$ PAH flux ratio depends on the size of the emitting PAHs \citep{Draine2001}. 
The non-detection of the $6.2\,\mu$m band sets a low upper limit of $R_{6.2/7.7} < 0.2$ on the PAH strength ratio,  that suggests preferentially large PAHs. Although the method used by \citet{Draine2001} to measure the PAH line strengths  is different from ours, we find that both $R_{7.7/11.3}$ and $R_{6.2/7.7}$ flux ratios in SQ can be explained by large (with a number of carbon atoms $N_{\rm c} \gtrsim 300$) and neutral PAHs in CNM conditions, excited with a $G_{\rm UV} \simeq 1$ interstellar radiation field \citep[see Fig.~17 of][]{Draine2001}. 



\section{Modeling dust emission}
\label{sec:models}

We present the physical framework and inputs of our modeling of the emission from dust associated with the molecular gas. 
Section~\ref{subsec:DUSTEM-PDR} and \ref{subsec:excitingfields} present the codes and the radiation field we use for our calculation of the dust emission.

\subsection{The DUSTEM (Dust Emission) code}
\label{subsec:DUSTEM-PDR}

We use an updated version of the \citet{Desert1990} model, the \textit{DUSTEM} code, to compute the dust emission.  The modifications\footnote{In particular, the absorption cross sections of the PAHs (with addition of new Aromatic Infrared Bands, AIBs) and the Big Grains (BGs), as well as the heat capacities (graphite, PAH C-H, silicate and amorphous carbon) have been updated.} implemented to the Desert et al. model are described in  \citet{Compiegne2008}. 
The dust properties, the dust-to-gas mass ratio, and the incident radiation field being given, the code calculates the dust SED $\nu \, S_{\nu}$ in units of erg~s$^{-1}$~H$^{-1}$, for each dust grain species, as a function of  the wavelength. 

We use the diffuse Galactic ISM size distribution \citep[a power-law  $n(a) \propto a^{-3.5}$, ][]{Mathis1977}, and dust-to-gas mass ratio infered from the fitting of the SED and extinction curve of the diffuse ISM \citep{Compiegne2008}.
The \textit{DUSTEM} model includes a mixture of three populations of dust grains with increasing sizes:
\begin{itemize}
\item Polycyclic Aromatic Hydrocarbons (PAHs) of radius $a = 0.4 - 1.2$~nm, responsible for the Aromatic Infrared Bands (AIBs) and the FUV non-linear rise in the extinction curve.
\item Very Small Grains (VSGs, $a = 1 - 4$~nm), which are carbonaceous (graphitic) nanoparticles producing the mid-IR continuum
emission and the extinction bump at 2175~\AA.
\item Big Grains (BGs, $a = 4 - 110$~nm) of silicates with carbonaceous
mantles or inclusions, which account for the far IR emission and the $1/\lambda$ rise at visible and near-IR wavelengths.
\end{itemize}

\subsection{Radiation field}
\label{subsec:excitingfields}

In this section we model the radiation field used to compute the dust emission. 
Stellar radiation is coming from the surrounding galaxies and/or stars in the ridge \citep{Gallagher2001}. The presence of ionizing radiation in the SQ ridge is also indicated by the lack of H{\sc i} gas in the H$_2$-bright shock structure \citep{Sulentic2001}, and by emission from ionized gas lines. Optical line emission diagnostics  suggest that  shocks  are responsible for hydrogen ionization in the ridge  \citep{Xu2003}. 


Therefore, we consider that the SED of the input radiation field consists of two components: a stellar component (sect.~\ref{subsubsec:ISRF}), and  photo-ionizing radiation from gas shocked at high-velocities (sect.~\ref{subsubsec:shockfield}). 

\subsubsection{The interstellar radiation field (ISRF)} 
\label{subsubsec:ISRF}

 The photometry performed on NUV and FUV \textit{GALEX} images  shows that UV flux in the shock, outside star-forming regions, corresponds to an interstellar radiation field of intensity $G_{\rm UV}=1.4 \pm 0.2$ (see  Table~\ref{table_UVfluxes}). 
What is the origin to this UV radiation in the shock? We estimate the UV emission in the ridge that comes from H{\sc ii} regions associated with star formation in the surrounding sources, i.e. NGC~7318a/b, the star-forming region SQ-A  at the northern tip of the shock structure, and NGC~7319. 
The extinction-corrected FUV luminosities of these main sources that surround the ridge are $L_{\rm FUV}(\rm NGC\,7318ab) = 8.1 \times 10^{9}\,$L$_{\odot}$, $L_{\rm FUV}(\mbox{SQ-A}) = 6.9\times 10^{9}\,$L$_{\odot}$, and $L_{\rm FUV}(\rm NGC\,7319) = 6.9\times 10^{9}\,$L$_{\odot}$ \citep{Xu2005}. Assuming that their distances to the center of the ridge are 10, 20 and 25~kpc, respectively, we find that the total UV flux in the ridge is $3.5 \times 10^{-3}$~erg~s$^{-1}$~cm$^{-2}$, i.e. $G_{\rm UV} \simeq 1.5$ in Habing units. 
Most of the UV flux comes from NGC~7318a/b. Although this simple calculation does not take into account the extinction between the UV sources and the ridge, it shows that this estimate of  the UV field is roughly consistent with our estimate from UV observations.  The very small number (of the order of unity) of star clusters that lie within the center of the SQ shock  \citep{Gallagher2001} shows that their contribution to the UV field is small. 


The SED of the stellar component of the radiation field is shown on Fig.~\ref{Fig_MIII_ISRF_inputPXDR_UV_NIR} (green dashed line). We use the \citet{Mathis1983} ISRF, scaled by a factor 1.4  to fit the FUV and NUV \textit{GALEX} photometry.  
The \textit{HST} $V$-band and \textit{WIRC} near-IR photometric measurements (see sect.~\ref{subsec:nearIRimaging} for observational details) are overlaid on Fig.~\ref{Fig_MIII_ISRF_inputPXDR_UV_NIR} (see  Table~\ref{table_2MASSfluxes}).  
The colors of the J, H and K$_{\rm s}$ fluxes do not exactly match that of the  \citep{Mathis1983} radiation field, but we remind that these measurements are uncertain (because at these wavelengths the radiation field is anisotropic, see sect.~\ref{photometry_method_radiation-field}).
This effect does not affect significantly our dust modeling since the near-IR part of the radiation field has a small impact on the dust emission. 


\subsubsection{Photo-ionizing radiation field: shock and precursor components} 
\label{subsubsec:shockfield}

  \begin{figure*}
    \centering
     \includegraphics[angle=90, width =  0.8\textwidth]{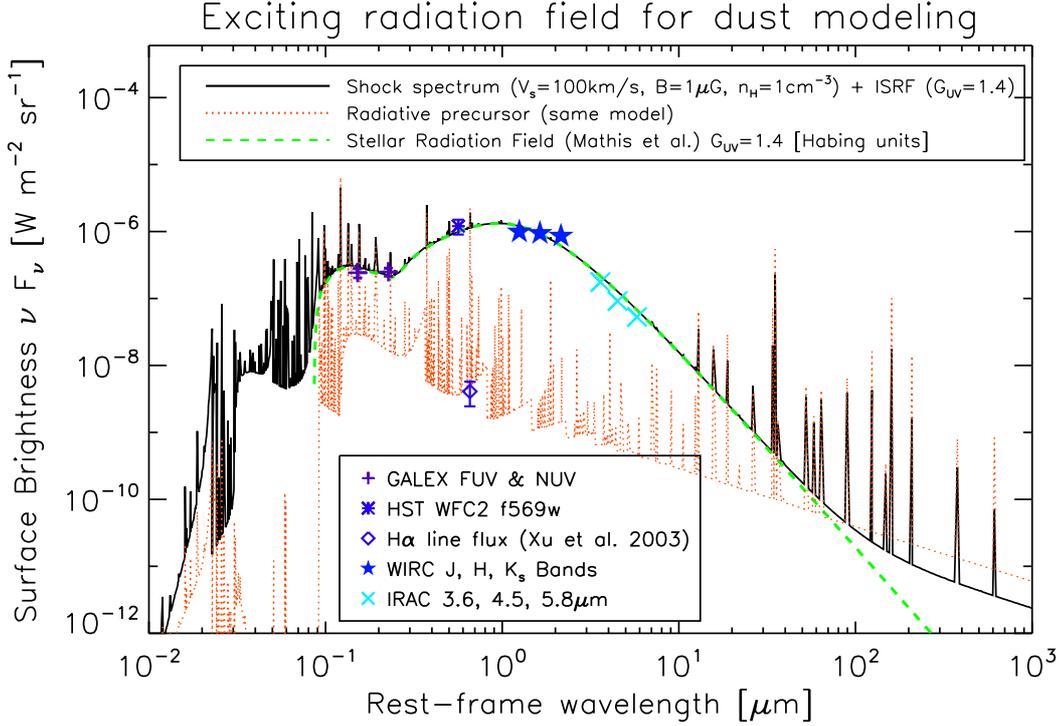}
     \caption{The 2-component SED of the  radiation field (stellar + shock) used as input for dust models to calculate the dust emission from molecular gas. The radiation field is the sum of  an ISRF of intensity $G_{\rm UV}=1.4$, scaled to match the observed UV flux in the shock, and the photo-ionizing spectrum of a $V_{\rm s} = 100$~km~s$^{-1}$ shock, scaled to match the H$\alpha$ emission in the SQ ridge. The preshock gas density is $n_{\rm H}= 1$~cm$^{-3}$ and the preshock magnetic field strength is $B_0 = 1 \,\mu$G.}
     \label{Fig_MIII_ISRF_inputPXDR_UV_NIR}
  \end{figure*}

In the shock sub-region, the [O{\sc i}]$6300\,$\AA~and [N{\sc ii}]$6584\,$\AA~to H$\alpha$ line ratios are $\sim 0.7$ and $\sim 0.3$, respectively, which is evidence of shock ionization \citep{Xu2003}. 
Therefore we model the SED of the ionizing field using emission from a radiative shock. In the following we constrain the shock parameters we use to model the emission from the ionized gas. 
The mid-IR \textit{IRS} spectrum extracted in the core of the shock (Fig.~\ref{Fig_IRS_spec_SQ_shock}) shows fine-structure line emission from [Ne$\,${\sc ii}]$\lambda \, 12.81\,\mu$m,  [Ne$\,${\sc iii}]$\lambda \,  15.56\,\mu$m, [Fe$\,${\sc ii}]$\lambda \,  25.99\,\mu$m, [S$\,${\sc iii}]$\lambda  \, 33.48\,\mu$m, and [Si$\,${\sc ii}]$\lambda  \, 34.82\,\mu$m. The spatial distribution of this emission and mid-IR line diagnostics of the excitation mechanisms are discussed in \citet{Cluver2010}. We summarize here the main results that are relevant to constrain the shock parameters. 

The comparison between the high value of $\mathcal{F}_{\rm [Ne II]12.8}  / \mathcal{F}_{\rm [Ne III]15.6} \sim 3.55$ and shock models \citep{Hartigan1987, Allen2008} allows to firmly constrain the range of shock velocities to $\sim 80 - 200$~km~s$^{-1}$  \citep{Cluver2010}. In the case of clumpy gas, if we consider the line emission from the shock only (discarding line emission from the pre-shock gas  ionized by emission from the shocked gas), the upper limit on shock velocities is a little higher (300 instead of 200~km~s$^{-1}$). 
In addition, comparison of the 
[S$\,${\sc iii}]$\lambda  \, 33.5\,\mu\rm m/$[S$\,${\sc iii}]$\lambda  \, 18.7\,\mu$m
 line ratios with these shock models constrains the pre-shock gas density to be $n_{\rm H} \lesssim 10\,$cm$^{-3}$. 
Therefore,  we adopt in the following a shock velocity of $V_{\rm s} = 100$~km~s$^{-1}$ and a pre-shock density of $n_{\rm H}= 1$~cm$^{-3}$. 
The photo-ionizing emission spectrum from shocked gas is taken from the library of the Mappings III shock code\footnote{\url{http://cdsweb.u-strasbg.fr/~allen/mappings_page1.html}} \citep{Allen2008}. We normalize the shock spectrum to  the observed H$\alpha$ flux\footnote{$\mathcal{F}_{\rm H\alpha} = 4.1  \times 10^{-9}$~W~m$^{-2}$~sr$^{-1}$ from \citet{Xu2003}.}. After this normalization, the SED of the ionizing shock emission is weakly sensitive to the two main shock parameters, the shock velocity and the gas density.


Fig.~\ref{Fig_MIII_ISRF_inputPXDR_UV_NIR} shows the SED of the radiation field we use as  input for the \textit{DUSTEM} code to compute the outcoming dust emission from the molecular and ionized gas. The black solid line is the  sum  of the two contributions (stellar + shock) to the radiation field. The shock spectrum itself is mostly composed of thermal bremsstrahlung (free-free) continuum and resonance lines arising from many different elements and ionic stages. It also shows a prominent low-temperature bound-free continuum of hydrogen, produced in the cool, partially-ionized zone of the recombination region of the shock, and the strong hydrogen two-photon continuum produced mostly by the down-conversion of Ly$\alpha$ photons trapped in this same region of the shock structure. Also present, though to a much weaker scale, is the bound-free continuum arising from the heavier elements, with the helium continuum the most obvious.

In the case of fast shocks ($V_{\rm s} > 100$~km~s$^{-1}$), the shocked medium emits UV radiation that ionizes the pre-shock medium before it is shocked. A so-called \textit{radiative precursor} propagates ahead of the shock, with an ionization front velocity that exceeds that of the shock. The contribution of the precursor itself is indicated on Fig.~\ref{Fig_MIII_ISRF_inputPXDR_UV_NIR} by the red dotted line. 
The green dashed line shows the contribution of the $G_{\rm UV}=1.4$ ISRF. The UV \textit{GALEX}, H$\alpha$ line, and near-IR \textit{WIRC} fluxes in the center of the shock are indicated.

The contribution of the precursor to the total H$\,${\sc ii} column density and radiative flux depends on the clumpiness of the pre-shock medium.
This contribution is negligible if the molecular gas is clumpy, i.e. fragmented into dense ($n_{\rm H} > 10^3$~cm$^{-3}$) clouds that have a small volume filling factor, because most of the ionizing photons do not interact with neutral gas but with the volume-filling, hot plasma that is optically thin to ionizing radiation. However, diffuse H$_2$ gas is expected to have a much higher volume filling factor. Our modeling of the  dust emission from the ionized gas takes into account the precursor contribution, weighted by the volume filling factor of the clumpy molecular gas.

\section{On the structure of the molecular gas}
\label{sec:results}

The dust emission depends on the optical thickness of the molecular gas to UV radiation. Is the molecular gas diffuse or fragmented in optically thick clumps?   
In section~\ref{subsec:structure_molgas} we describe the assumptions we make regarding the structure of the molecular gas. 
Then we present the results of our modeling of the dust emission in the SQ shock.  We discuss the dust emission for the two physical structures of the H$_2$ gas presented in sect.~\ref{subsec:structure_molgas}: diffuse  (sect.~\ref{subsubsec:model_op_thin}) and clumpy molecular gas (sect.~\ref{subsubsec:model_op_thick}). 
Both models include the contribution of the ionized gas (H$\,${\sc ii}) to the dust emission (this calculation is detailed in sect.~\ref{dust_em_HIIgas}).
This detailed modeling is ﬁtted to mid-IR observations, and used to investigate the inﬂuence of the 
structure of the molecular gas on the FIR dust SED.

\subsection{Two limiting cases for the structure of the molecular gas}
\label{subsec:structure_molgas}


  \begin{figure}
      \includegraphics[width = 0.24 \textwidth]{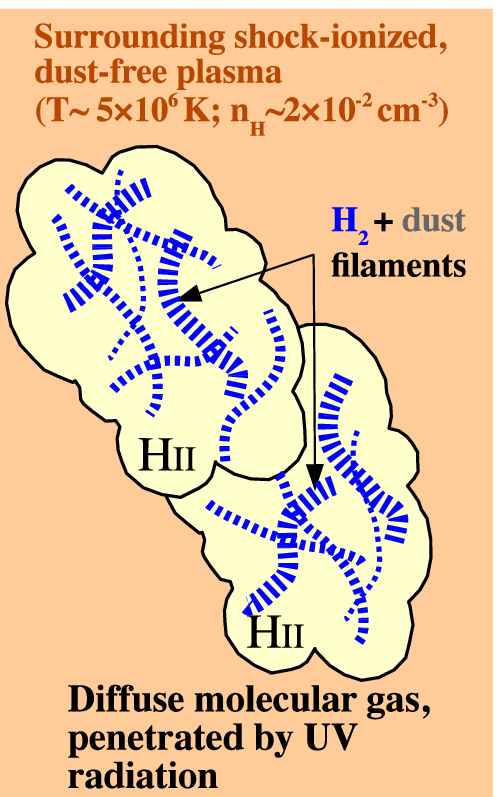}
      \includegraphics[width = 0.24 \textwidth]{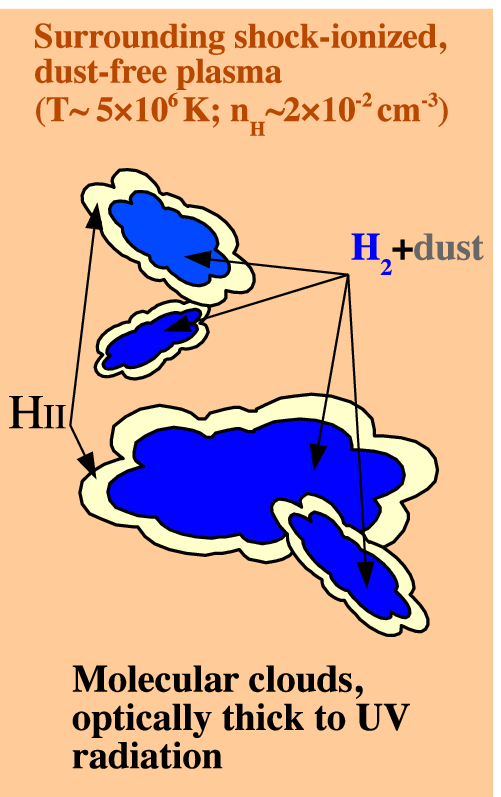}
      \caption{Sketch of the two limiting physical states of the multiphase molecular gas we are considering in the framework of our modeling of dust emission. \textit{(Left)}  Dust is associated with \textit{diffuse} molecular gas that  is broken into fragments, filaments or sheets, penetrated by UV radiation. \textit{(Right)} The dusty molecular gas is in clouds that are \textit{optically thick} to UV light. }
     \label{Fig_mol_gas_state}
 \end{figure}

The spectral energy distribution of the dust emission depends also on the structure of the molecular gas, and in particular on the optical depth of the clouds. 
The physical structure of the molecular gas in the SQ shock is an open question. 
In the following we explore these two cases.
\begin{enumerate}
\item The molecular gas is \textit{diffuse} as the gas observed in the solar neighborhood through UV spectroscopy \citep[e.g.][]{Rachford2002} and across the Galaxy through observations of mid-IR H$_2$ line emission \citep{Falgarone2005}. In this case, we assume that the molecular gas is optically thin to UV radiation.
\item The molecular gas is within \textit{clouds} that are \textit{optically thick} to UV photons, as those observed in star-forming regions.
\end{enumerate}
Fig.~\ref{Fig_mol_gas_state} illustrates, in a simplistic manner, the two physical states (diffuse or clumpy) of the molecular gas we are considering. In both cases, the molecular gas is embedded within H{\sc ii} gas and shock-heated, X-ray emiting plasma. 

In order to compute the dust emission from optically thick (to UV photons) clouds (Fig.~\ref{Fig_mol_gas_state}, right panel), we use the \textit{Meudon PDR} (Photon Dominated Region) code \citep{LePetit2006}
  to compute the radiative transfer through the cloud. This 1-dimensional, steady-state model considers a stationary plane-parallel slab of gas and dust, illuminated by UV radiation. 
The radiation field output $I_{\nu}$ of the Meudon PDR code is used as input to
the \textit{DUSTEM }program to compute the spectral energy distribution (SED) of the
dust as a function of the optical depth into the cloud. This 2-step process is iterated to take into account dust heating by the dust IR emission. Usually $4-5$ iterations are needed so that the radiation field converges.

\subsection{Dust emission from diffuse molecular gas}
\label{subsubsec:model_op_thin}

  \begin{figure*}
     \centering
     \includegraphics[width =  0.9\textwidth]{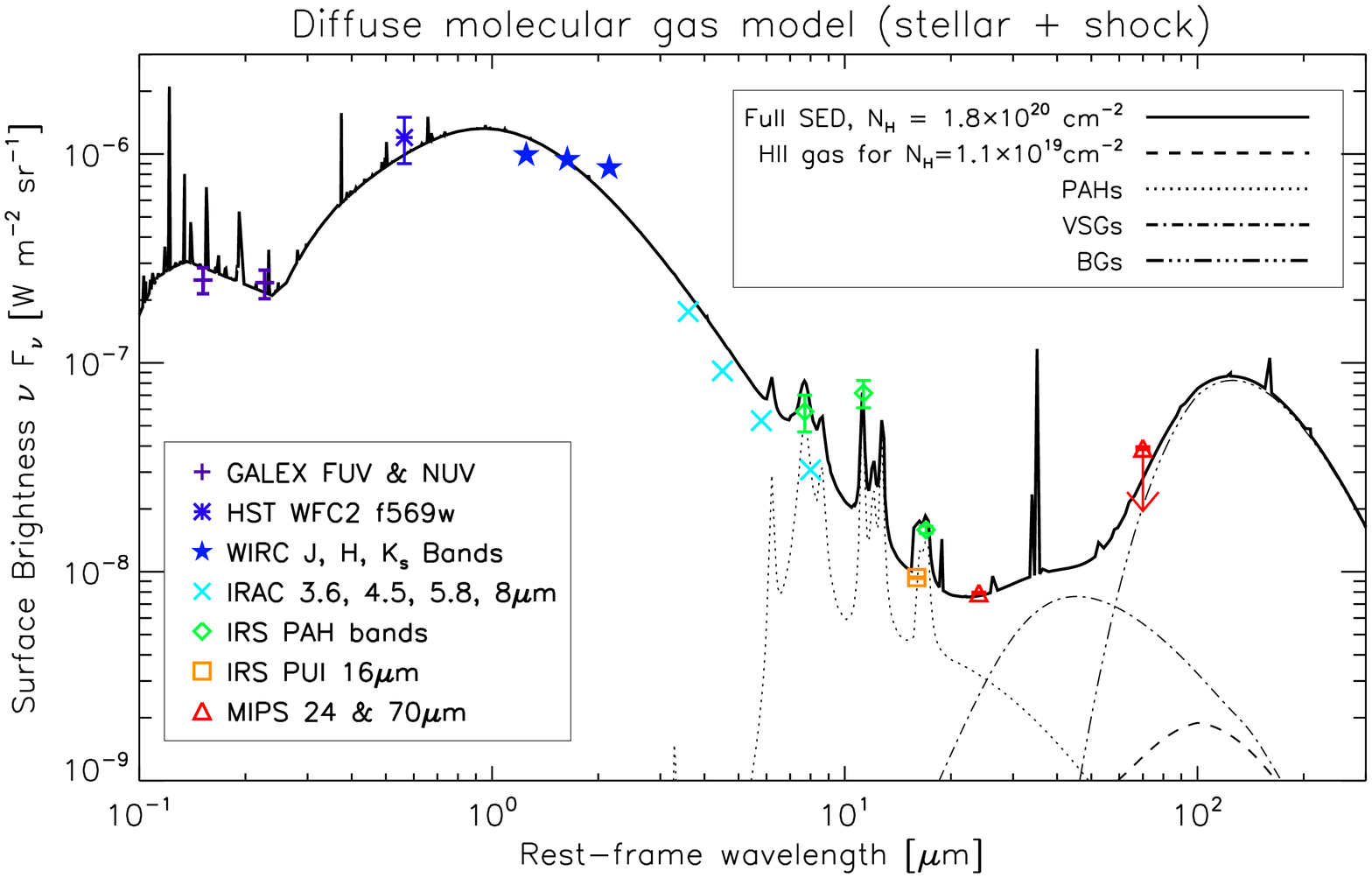}
     \includegraphics[width =  0.9\textwidth]{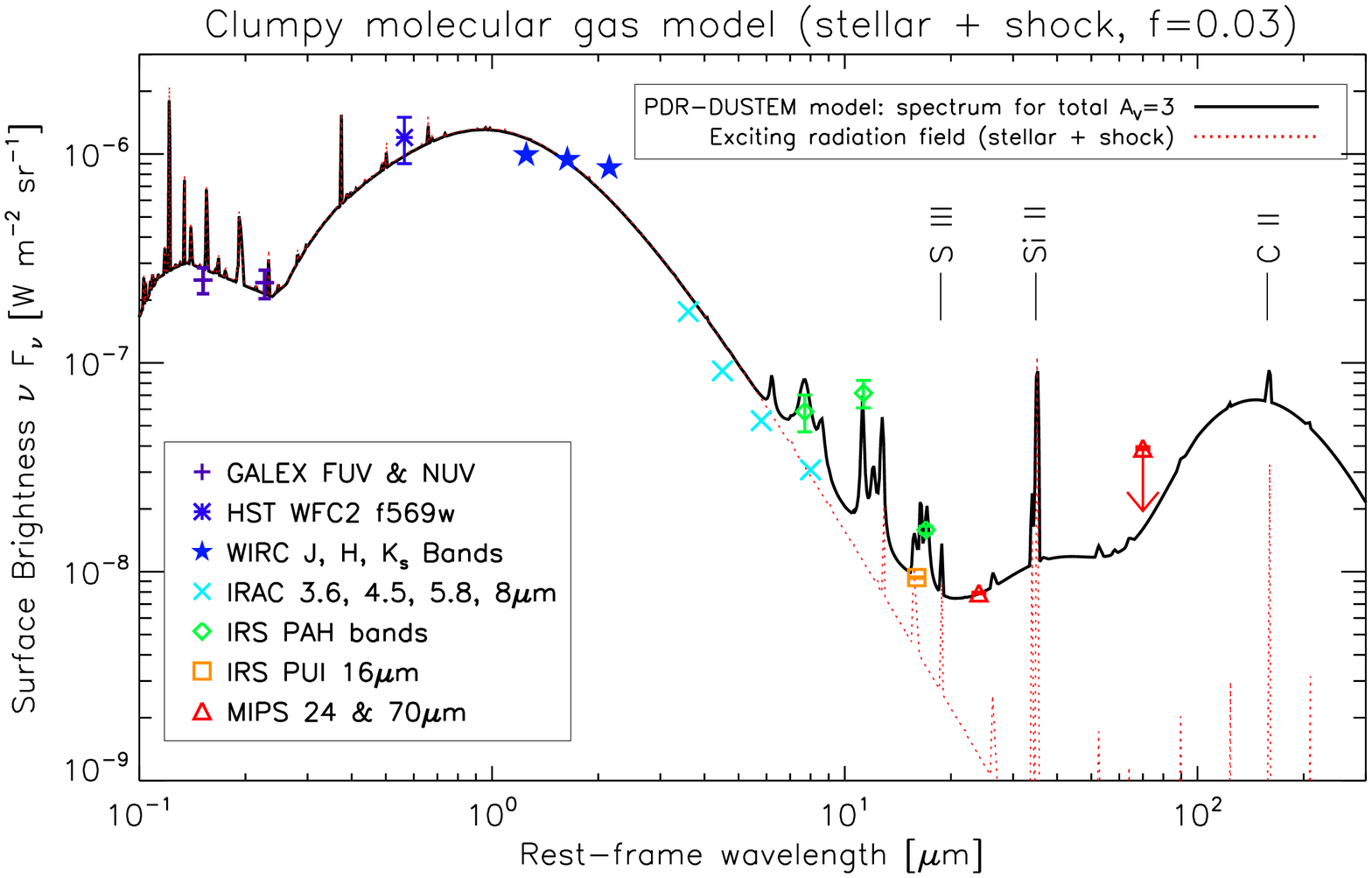}
     \caption{SED of the dust emission associated with diffuse \textit{(top)} or clumpy  \textit{(bottom)} molecular gas. The cloud is exposed to a radiation field consisting of a mixture (stellar + shock) of the \citet{Mathis1977} ISRF (scaled by a factor $G_{\rm UV}=1.4$) and the ionizing emission spectrum of shock-heated gas for a shock velocity of 100~km~s$^{-1}$. \textit{(top)} The dashed line shows the contribution of the ionized gas (see text for details). 
The other broken lines show the different contributions of the three populations of grains (PAHs, VSGs, and BGs). We do not show them on the bottom figure for clarity. Assuming a Galactic dust-to-gas mass ratio, a column density of $N_H = 1.8 \pm 0.5 \times 10^{20}$~cm$^{-2}$ is needed to fit the \textit{Spitzer} data. 
\textit{(bottom)} The black solid line is the sum of the incoming and outcoming emission from a slab of gas of total $A_{\rm V} = 3$, calculated with the \textit{Meudon PDR} and \textit{DUSTEM} codes. To fit the \textit{Spitzer} data, the model IR emission is scaled by a factor $f= 0.03$, which represents the effective surface filling factor of the molecular gas. The observed IRAC and MIPS fluxes correspond to the entries of  Table 1. The green points are the IRS fluxes at the peak of the 7.7, 11.3 and 17~$\mu m$ PAH bands.
}
     \label{Fig_dust_spec_Go1_4}
\label{Fig_dust_spec_PXDR_MIII_Vs100}
  \end{figure*}

The \textit{DUSTEM} code is used to model the dust emission from diffuse molecular gas (sect.~\ref{subsec:DUSTEM-PDR}), penetrated by UV photons.
The cloud is illuminated by a composite field, made up of a stellar component (ISRF of intensity $G_{\rm UV} = 1.4$ Habing units) and the ionizing radiation from the shocked gas (see sect.~\ref{subsec:excitingfields} and Fig.~\ref{Fig_MIII_ISRF_inputPXDR_UV_NIR}). 
The dust properties are those of sect.~\ref{subsec:DUSTEM-PDR}. The abundances and size distribution of the dust grains are Galactic. 
The dust emission from the ionized gas is included in the model SED. It is calculated separately with the \textit{DUSTEM} code and added to the contribution of the cloud (see sect.~\ref{dust_em_HIIgas}). 

The top panel of Fig.~\ref{Fig_dust_spec_Go1_4} shows the $0.1 - 300\,\mu$m SED resulting from the model (black solid line), so that both the UV radiation field and IR dust emission can be seen. 
The dashed line (in the bottom right corner of the plot) shows the dust emission from the H$\,${\sc ii} gas. 
The three other broken curves show the contributions of the different populations of dust grains, i.e. PAHs, VSGs and BGs, respectively.
The \textit{GALEX}, \textit{WIRC} and \textit{Spitzer} fluxes are indicated for comparison (see  sect.~\ref{sec:photometry}, Tables~\ref{table_midIR_fluxes}, \ref{table_UVfluxes} and \ref{table_2MASSfluxes}). 
The 7.7, 11.3 and $17\,\mu$m points are the peak values of the PAH bands detected in the \textit{IRS} spectrum (sect.~\ref{subsec:IRS}). Other points come from imaging broadband photometry (GALEX, HST, IRAC and MIPS) measurements performed in the center of the SQ ridge (over an aperture of $17''$ in diameter, see sect.~\ref{sec:photometry}). 
%

The 11.3, 16 and $24\,\mu$m data points are used to fit the model SED. As it may be a lower limit, the $70\,\mu$m point has not been included in the fit.
The gas column density is the only free parameter to fit the data. 
The model SED is obtained by multiplying the emissivity output of the \textit{DUSTEM} code by the column density of warm molecular gas that is determined from fitting the SED to the mid-IR data. 
 We find that the column density that best fits the data is  
 $N_{\rm H} = 1.8 \pm 0.5 \times 10^{20}$~cm$^{-2}$ for a Galactic dust-to-gas mass ratio.  
This column density obtained from the model is close to the column density derived from the H$_2$ line fluxes ($N_{\rm H} \simeq 2.2 \times 10^{20}$~cm$^{-2}$) within this aperture. This column density is obtained by fitting the rotational H$_2$ line fluxes with C-shocks models as described in \citet{Guillard2009}.
This suggests that the column density  of cold ($T \lesssim 50$~K) molecular gas is much smaller than that of the warm H$_2$. However, this may not be the case because the dust-to-gas mass ratio is possibly lower than the Galactic value we have assumed here,  and the molecular gas may be in clumps optically thick to UV photons (see next section). 

\subsection{Dust emission from clumpy molecular gas}
\label{subsubsec:model_op_thick}

We model the dust emission associated with molecular gas fragmented into clumps that are optically thick to UV radiation.
Here the \textit{DUSTEM} and \textit{Meudon PDR} codes (see sect.~\ref{subsec:DUSTEM-PDR}) are combined to calculate the emission from a molecular cloud of total $A_{\rm V} = 3$. The column density of the cloud is thus $N_{\rm H}^{\rm (c)} = 6.9 \times 10^{21}$~cm$^{-2}$. The choice of the $A_{\rm V} $ value is  arbitrary.
We check that the model SED in the mid-IR domain does not depend much on the total $A_{\rm V}$ because the mid-IR emission comes mainly from the surface of the cloud. 
The cloud is illuminated by a composite field (stellar + shock), made up of the ISRF of intensity $G_{\rm UV} = 1.4$ (Habing units) and the ionizing radiation from the shocked gas (see sect.~\ref{subsec:excitingfields} and Fig.~\ref{Fig_MIII_ISRF_inputPXDR_UV_NIR}). 

Fig.~\ref{Fig_dust_spec_PXDR_MIII_Vs100} shows the sum of the outcoming radiation (cloud + ionized gas) emission spectrum, plus the incoming radiation, from the UV to the FIR ($0.1 - 300\;\mu$m). The dashed line shows the emission from the ionized gas. The UV, and IR photometric measurements are overlaid for comparison between data and models.

From UV to FIR wavelengths, the total SED (solid black line) consists of the ISRF, including the free-free, free-bound and resonance line emission from the shocked gas, and the dust emission including  the AIBs, the emission from the VSGs and the grey body of the BGs at long wavelengths. The spectrum also shows some fine-structure IR lines, indicated on the spectrum, e.g. $18.7\,\mu$m [S {\sc iii}], $34.8\,\mu$m [Si {\sc ii}], and $157.7\,\mu$m [C {\sc ii}].
To fit the data, the model IR emission from the cloud is scaled by a factor of $f=0.03$, which represents an effective surface filling factor for the molecular gas.
The low value of $f$ is consistent with the low value of the average optical extinction in the shock region.
 For clumps of $A_{\rm V} = 3$, the average column density of the cloud phase is thus $\langle N_{\rm H}  \rangle = f \times N_{\rm H}^{\rm (c)} \simeq 2.1 \times 10^{20}$~cm$^{-2}$. 
Our specific choice of $A_{\rm V} = 3$ leads to a value of  $\langle N_{\rm H}  \rangle$ that is consistent with the column density of the warm H$_2$ gas seen by \textit{Spitzer}. For larger  values of   $A_{\rm V}$, the total column density of molecular gas will be larger  than the column density of warm H$_2$ inferred from rotational lines. 



\subsection{Dust emission from shock- and precursor-ionized gas}
\label{dust_em_HIIgas}

We detail how we calculate the emission from dust associated with the ionized gas. This contribution is added to both models (diffuse and clumpy). We use the input radiation field of Fig.~\ref{Fig_MIII_ISRF_inputPXDR_UV_NIR}, including the ionizing part of the spectrum ($\lambda < 912$~\AA), to compute the dust emissivity with the \textit{DUSTEM} code. The column density of ionized gas has been constrained by the integrated flux of the $12.8\,\mu$m [Ne$\,${\sc ii}] fine-structure line, $\mathcal{F}_{\rm [Ne\,II]}$, measured on the \textit{IRS} spectrum. The $\mathcal{F}_{\rm [Ne\,III]} / \mathcal{F}_{\rm [Ne\,II]}$ line flux ratio is low\footnote{$\mathcal{F}_{\rm [Ne\,III]} / \mathcal{F}_{\rm [Ne\,II]} = 0.14 \pm 0.04$ in the center of the SQ ridge, see Table~3 in  \citet{Cluver2010}.},  which implies that Ne$\,${\sc ii} is the main ionization state of Ne in the shock. Therefore, the column density of ionized gas can be expressed as:
\begin{equation}
\label{eq:column-density-NeII}
N_{\rm H}^{\rm (ion. gas)}  = 2.3 \times 10^{19} \   \frac{\mathcal{F}_{\rm [Ne\,II]} \  \rm [W\,m^{-2}\,sr^{-1}]}{1.17 \times 10^{-9}} \, \frac{10 \ \rm cm^{-3}}{n_e} \  \ \rm cm^{-2} \ .
\end{equation}
We use the relation between the [Ne$\,${\sc ii}] line intensity and the emission measure given in \citet{Ho2007} and we assume that the electronic density of the $10^{4}$~K gas is $n_e = 10$~cm$^{-3}$. This value is consistent with the SQ postshock pressure ($\sim 2 \times 10^5$~K~cm$^{-3}$). The column density derived from Eq.~\ref{eq:column-density-NeII} is comparable to the H$\,${\sc ii} column density derived from the shock model ($N_{\rm H}$(H{\sc ii})$ = 1.2 \times 10^{19}$~cm$^{-2}$ for a shock + precursor model at a shock velocity of 100~km~s$^{-1}$ and a preshock density $n_{\rm H} = 1$~cm$^{-3}$).


The dust emission associated with this amount of ionized gas is the black dashed line on the top plot of Fig.~\ref{Fig_dust_spec_PXDR_MIII_Vs100}.
The column density of the ionized gas being one order of magnitude smaller than that of the warm H$_2$, the contribution of the ionized gas to the IR dust emission is  negligible.  

\subsection{Comparison between models and degeneracies}
\label{subsec_compModels}

  \begin{figure*}
\centering
     \includegraphics[width = \textwidth]{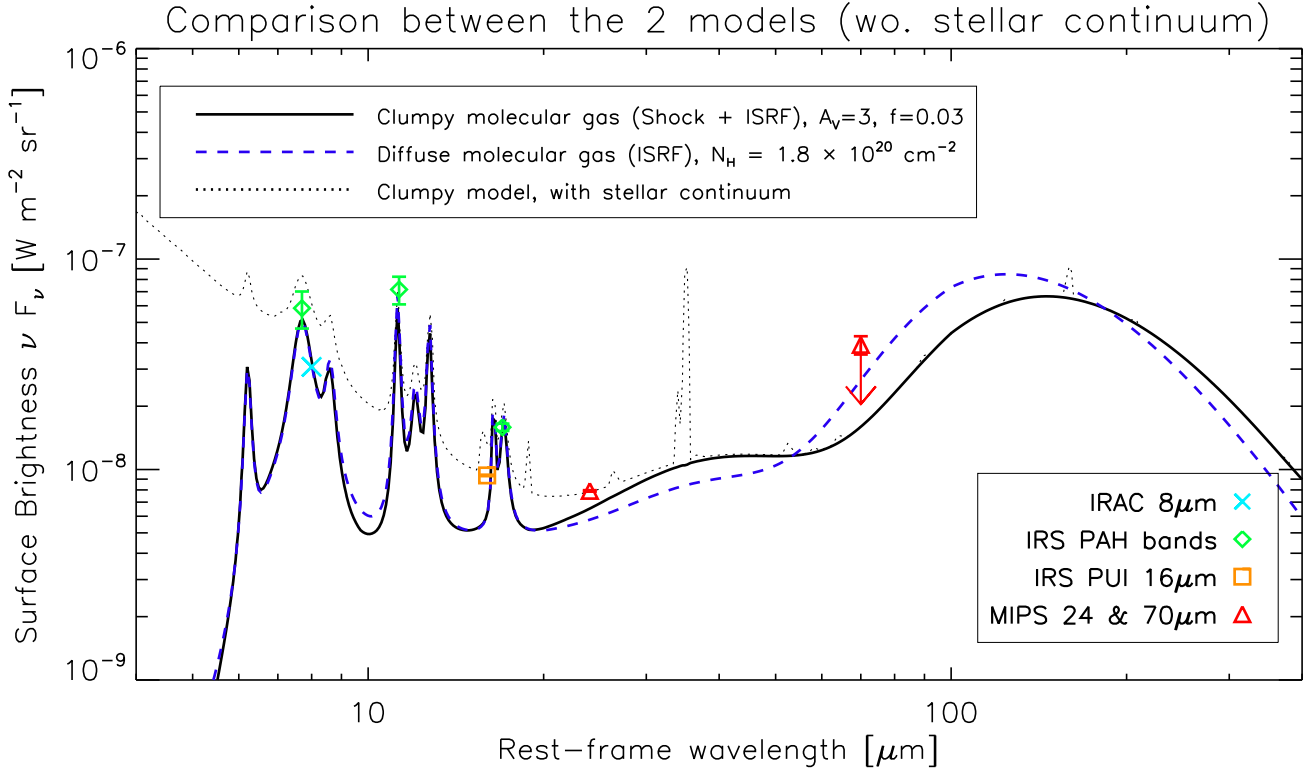}
     \caption{Comparison between the two models presented in this paper:  diffuse (dashed blue line, same model as Fig.~\ref{Fig_dust_spec_Go1_4}) and clumpy (black line, same model as Fig.~\ref{Fig_dust_spec_PXDR_MIII_Vs100}) molecular gas. The overlaid points indicate the \textit{Spitzer} fluxes extracted within the shock region (see bottom right inset).  
The gas lines are removed from these spectra, as well as the IR component of the incident ISRF ($G_{\rm UV}=1.4$).  For comparison, the clumpy model with stellar continuum is shown on the thin black dotted (same model as Fig.~\ref{Fig_dust_spec_PXDR_MIII_Vs100}). 
}
     \label{Fig_dust_spec_comp}
  \end{figure*}

Fig.~\ref{Fig_dust_spec_comp} shows the comparison between the result of the two models presented above.
We focus on the dust emission, so the gas lines and the stellar continuum from the ISRF are removed from the spectra.
The FIR SEDs are different between the two models. The FIR peak brightness of the SED of the diffuse molecular gas is brighter (by a factor of 1.7) and shifted towards shorter wavelengths compared with the clumpy  model. 
This difference can be easily explained. When the molecular gas is clumpy, the dust grains see, on average, an attenuated radiation field, and are thus colder than in the diffuse model. For a given column density of matter, the FIR brightness is also fainter.

Present observations do not allow us to decide whether the gas is diffuse or fragmented into optically thick clumps. The spatial resolution  of \textit{Spitzer} at long wavelengths ($\lambda \gtrsim 70\,\mu$m) is not high enough to obtain  accurate photometric measurements. FIR observations with the \textit{Herschel Space Telescope} will provide the sensitivity and angular resolution needed to test our models.
However, it may not be so straightforward to conclude on the structure of the molecular gas because we assume a Galactic dust-to-gas mass ratio and size distribution.
The IR SED also depends on the relative abundance between VSGs and BGs. 

We defer the discussion about future observations needed to 
disentangle the cloud structure and the dust size distribution to sect.~\ref{sec:conclusion}. Also note that we cannot estimate the total dust extinction from the models, given the lack of observational constrains in the FIR. The $A_{\rm V}$ value may be higher than the value derived from our modeling if the total H$_2$ column density including cold molecular gas is higher than 
 that derived from the mid-IR H2 rotational lines. If we assume that the molecular gas is diffuse and $N_{\rm H}=2\times 10^{20}$~cm$^{-2}$, this corresponds to $A_{\rm V}=0.1$. This is smaller than the average value derived from optical observations ($A_{\rm V} \sim 0.6$ in the main shock region for the diffuse emission). 

\section{Dust processing in the Stephan's Quintet shock?}
\label{sec:discussion}

The Stephan's Quintet galaxy-wide collision is an extreme environment where observations and modeling dust emission may provide insight into dust processing in shocks. 
The galaxy collision must have triggered shocks across the whole ISM. As discussed in \citet{Guillard2009}, the shock velocity depends on the preshock gas density. Gas at preshock densities  $n_{\rm H} > 0.2\,$cm$^{-3}$ has been shocked at velocities small enough ($V_{\rm s} < 200$~km~s$^{-1}$) to cool, to keep most of its dust, and to become molecular within a few million years.
To account for the H$_2$ emission, the molecular gas must be processed by low-velocity ($5-20$~km~s$^{-1}$) MHD shocks, repeatedly. Therefore, the origin and dynamical state of the SQ molecular gas is very different from that of the Galactic ISM. 


In the previous section, we assume that the dust properties in the SQ ridge are identical to those of the Galaxy, which obviously is a simplifying assumption, which may be far from reality. In  the shock region, various processes can affect dust grains   \citep[e.g. thermal sputtering of grains in the hot gas or destruction in shocks due to gas-grain and grain-grain interactions, see][and references therein]{Jones2004} that can affect both the dust-to-gas mass ratio and the dust size distribution. So far, observational evidence of dust processing comes mainly from gas phase metals depletions \citep{Sembach1996}. To our knowledge, there is no direct observational evidence for changes in the dust size distribution that can be unambiguously associated with shock-processing. 
Stephan's Quintet is an outstanding target to look for such evidence on galactic scales. 

Dust destruction processes depend on the type of shocks. 
 \citet{Guillard2009} show that within the timescale of the galaxy collision ($\sim 5 \times 10^{6}$~yr), the destruction of grains smaller than $\sim 0.1\,\mu$m is complete  in the hot,  teneous gas (corresponding to pre-shock densities $n_{\rm H} \lesssim \times 10^{-2}\,$cm$^{-3}$)  that is shocked at high velocities ($V_{\rm s} > 300$~km~s$^{-1}$). For intermediate preshock densities ($n_{\rm H} \sim 0.2 - 0.01\,$cm$^{-3}$, $V_{\rm s} = 100 - 300$~km~s$^{-1}$), models predict significant dust destruction ($10 - 50\,$\% in mass) and possibly the production of an excess of small grains by shattering \citep[e.g.][]{Jones1994}. 
Within the dense ($n_{\rm H} > 10^{3}\,$cm$^{-3}$) molecular gas, low-velocity MHD  shocks ($5-20$~km~s$^{1}$), may only have little effect on dust \citep[][]{Guillet2007, Gusdorf2008}.



The present SQ data provide some constraints on the dust size distribution.
The detection of PAH emission from such a violent and extreme environment as the SQ shock may be seen as a surprise.
 PAHs are predicted to be completely destroyed for $V_{\rm s} \gtrsim 125 $~km~s$^{-1}$ ($n_{\rm H} = 0.25\,$cm$^{-3}$) and their structure would be deeply affected for $  50 \lesssim V_{\rm s}  \lesssim 100 $~km~s$^{-1}$ \citep{Micelotta2010}. 
Therefore, the PAH detection provides interesting constraints on the density structure of the  preshock gas.
Since it is unlikely that PAHs reform efficiently from carbon atoms in the postshock gas, we conclude that \textit{(i)} they were protected from high-velocity shocks in high preshock density regions ($n_{\rm H} \gtrsim 0.3\,$cm$^{-3}$), or/and \textit{(ii)} they are the product of the shattering of VSGs in the shock.

We note that the $6.2\,\mu$m AIB is absent and that the $17\,\mu$m complex is prominent. This suggest that large PAHs, which emit more efficiently at larger wavelengths  \citep[e.g.][]{Draine2007}, would be predominant in the shock. This may result from PAH processing in shocks, larger molecules being less fragile than smaller ones \citep{Micelotta2010}. This interpretation is supported by \textit{Spitzer} observations by \citet{Tappe2006} that show a prominent $17\,\mu$m emission from the supernova remnant N132D in the Large Magellanic Cloud.

  \begin{figure}
     \centering
     \includegraphics[angle=90,width = \columnwidth]{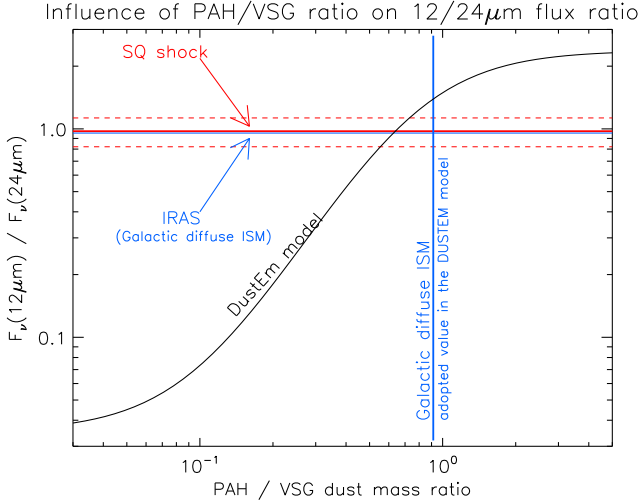}
 \caption{Predicted $12 / 24 \,\mu$m flux ratio from the \textit{DUSTEM} model as a function of the PAH / VSG mass ratio. The horizontal red and blue lines shows the observed $12 / 24 \,\mu$m flux ratios for the SQ shock, and the Galactic diffuse ISM (\textit{IRAS} observations). The vertical blue line is the  Galactic PAH/VSG ratio adopted in the model (0.915). This value is chosen to fit both the  $12 / 24 \,\mu$m ratio and the extinction curve.}
     \label{Fig_Ratio_12_24_PAH_VGS}
  \end{figure}

The $12 / 24\,\mu$m flux ratio is sensitive to the relative mass abundances between PAHs and VSGs. 
 Fig.~\ref{Fig_Ratio_12_24_PAH_VGS} shows the predicted $12 / 24\,\mu$m flux ratio as a function of the PAH / VSG dust mass ratio. For each PAH/VSG ratio, the dust emission spectrum is calculated for an ISRF of $G_{\rm UV} = 1.4$ with the \textit{DUSTEM} code. We derive the $12$ and $24\,\mu$m fluxes by integrating the model spectrum over the $12\,\mu$m \textit{IRAS} and $24\,\mu$m \textit{MIPS} filter bandpasses. 
The PAH/VSG mass ratio in the SQ shock is remarkably close  to the value for the Galactic diffuse ISM. A deviation would be expected if the dust had been processed by high-speed ($> 100$~km~s$^{-1}$) shocks \citep{Jones1996}. Like for PAHs, this may suggest that the postshock dust was lying in gas dense enough to have been protected
from destruction by fast shocks, and thus was protected from the effect of fast shocks, as proposed by \citet{Guillard2009}. However, as discussed in sect.~\ref{subsubsec:PAH}, the PAH spectrum in the shock is significantly different from that observed in the diffuse Galactic ISM. The fact that the long wavelength bands (11.3 and 17$\,\mu$m) are brighter than the 6.2 and 7.7$\,\mu$m features imply that a higher fraction of the PAH emission is emitted within the 12$\,\mu$m IRAS band. If true, this correction would imply a lower PAH/VSGs abundance than in the Galactic diffuse ISM.

Far-IR SED imaging of the dust, possible with the PACS\footnote{Photodetector Array Camera \& Spectrometer} and SPIRE\footnote{Spectral and Photometric Imaging Receiver} instruments onboard \textit{Herschel}, combined with modeling of the full IR spectral energy distribution, would provide diagnostics for measuring the relative abundances of the different grain populations. 




In our modeling (sect.~\ref{sec:results}), we did not consider any thermal emission from collisionally heated dust in the hot plasma.
This emission may arise after a fast shock wave has gone through tenuous, dusty gas, but it will last a very short period of time ($\sim 10^6$~yr), producing a ``flash'' of FIR emission, because the dust cooling efficiency drops as the grain sputtering occurs in the hot ($> 10^{6}$~K) gas \citep{Smith1996, Guillard2009}. However,  there are two reasons why this may not be a valid assumption. First, if there is a significant dust mass in grains larger than about $ 0.3\,\mu$m, this dust  may survive for $ \gtrsim 5 \times 10^{6}$~yr. These grains may contribute to the FIR emission, as proposed by \citet{Xu2003} and further investigated by \citet{Natale2010}. Second, we cannot exclude that some dust may be injected into the hot phase by the ablation of clouds (turbulent mixing), due to their dynamical interaction with the background plasma \citep{Guillard2009}. 
A continuous supply of dust from the warm to hot phase could balance destruction by sputtering. This possibility needs to be quantified, which is beyond the scope of this paper. 

%

\section{Summary and concluding remarks}
\label{sec:conclusion}

In this paper, we present new \textit{Spitzer} imaging and spectroscopic observations that reveal PAH and VSG emission from the galaxy-scale shock in Stephan's Quintet (SQ). 
Here are our main observational results:

\begin{itemize}
\item Faint dust emission is detected in the center of the H$_2$-bright SQ shock structure, outside star-forming regions lying in the SQ halo. 
The $12 / 24 \,\mu$m flux density ratio in the SQ ridge is remarkably similar to that of the diffuse Galactic ISM. 
This suggests that the PAH to VSG abundance ratio is similar to that of the diffuse ISM of the Galaxy.
\item 
The global  mid-IR SED is consistent with the expected dust emission from the amount of warm H$_2$ detected by \textit{Spitzer} ($N_{\rm H} \simeq 2 \times 10^{20}$~cm$^{-2}$) for a UV radiation field intensity of $G_{\rm UV} \sim 1$ [Habing unit], which is consistent with UV observations of the shock. 

\item The PAH emission spectrum in the SQ shock is significantly different from that of the diffuse Galactic ISM. The $7.7$, $11.3$ and $17\,\mu$m aromatic bands are detected, but the $6.2\,\mu$m band is absent. The $17\,\mu$m complex is prominent, but the $16.4\,\mu$m is not detected. Interestingly,  the $7.7 / 11.3\,\mu$m flux ratio in the SQ shock  is a factor $\sim 2$ lower than that of the diffuse Galactic PAHs. These characteristics  may suggest an enhanced fraction of neutral and large PAHs.
\end{itemize}

\textit{Spitzer} imaging and spectroscopy reveal powerful H$_2$ emission in the Stephan's Quintet X-ray giant shock that extends over the full area ($\approx 35 \times 15$~kpc$^2$) of the ridge \citep{Cluver2010}. 
We expect some dust emission to come from molecular gas because H$_2$ forms on dust grains. 
In this paper we test this interpretation by modeling the IR emission from dust associated with the H$_2$  gas present in the SQ shock structure, and by comparing the model calculations with  \textit{Spitzer} observations.

\begin{itemize}
\item We model the emission from dust associated with diffuse or clumpy molecular gas, embedded within H{\sc ii} gas and X-ray emitting plasma. The model SED is consistent with mid-IR \textit{Spitzer} observations for both cases, for a Galactic dust-to-gas mass ratio and a Galactic dust size distribution. For diffuse gas, the best-fit column density is $N_{\rm H} = 1.8 \pm 0.5 \times 10^{20}$~cm$^{-2}$, which is close to the value derived from warm H$_2$ observations. For clumpy molecular clouds that are optically thick to UV radiation, we find that the H$_2$ surface filling factor is $f \sim 0.03 $.
So far, the present data and the degeneracy between the dust size distribution and the cloud structure do not allow us to decide whether the molecular gas is diffuse or fragmented into clouds that are optically thick to UV photons. 

\item The presence of dust in the SQ shock shows that dust is able to survive in such a violent environment. We propose that at the time of the high-speed galaxy encounter, the dust that survived destruction was in pre-shock gas 
at densites larger than a few $0.1\,$cm$^{-3}$, which was not shocked at velocities larger than $\sim \!200\,$km~s$^{-1}$ \citep[see ][]{Guillard2009}. Present data do not allow us to indentify a plausible impact of the shock  
on the dust size distribution, but the peculiar properties of the 
PAH emission in the SQ ridge (summarized above) may be the result of PAH processing in shocks.

\end{itemize}


Future Far-IR observations are needed to constrain both the structure of the H$_2$ gas and the dust properties in the SQ shock. The PACS and SPIRE instruments on \textit{Herschel} would provide SED of the Far-IR dust emission
resolved on scales of $5-10$~arcsecs, crucial  for helping to distinguish how clumpy the dust distribution is in the shock structure. 
The flux in the FIR is different for the two models (diffuse or clumpy, see Fig.~\ref{Fig_dust_spec_comp}). The spatial resolution of the PACS [$60-210 \, \mu$m] instrument on-board the \textit{Herschel Space Observatory} at $70\,\mu$m is comparable to that of \textit{Spitzer} at 24~$\mu$m. It will thus be possible to extract the FIR emission from the shock with much more accuracy than with  \textit{Spitzer}. This is needed to possibly break some degeneracies of the model and better constrain the key-parameters (e.g. dust size distribution).


The Stephan's Quintet galaxy-collision is a unique environment to study dust survival in shocks,  on galactic scales. The presence of dust in the SQ shock stresses the need to revisit  the standard calculations of dust survival timescales in the ISM, by taking into account its multiphase structure.

\begin{acknowledgements}
This work is partly based on observations made with the Spitzer Space Telescope, which is operated
by the Jet Propulsion Laboratory, California Institute of Technology under a contract
with NASA. 

GALEX (Galaxy Evolution Explorer) is a NASA small explorer
launched in 2003 April. We gratefully acknowledge
NASA's support for construction, operation, and science analysis
for the GALEX mission, developed in cooperation with the
Centre National d'Etudes Spatiales (CNES) of France and the Korean
Ministry of Science and Technology.

This research has made use of the NASA/IPAC Extragalactic Database (NED)
which is operated by the Jet Propulsion Laboratory, California Institute of Technology, under
contract with the National Aeronautics and Space Administration.


PG would like to thank M. Gonzalez Garcia and J. Le Bourlot for help about the PDR code, and V. Guillet for helpful discussions about dust processing in shocks. We wish to aknowledge R. Tuffs, C. Popescu, G. Natale and E. Dwek for fruitful discussions we had about dust emission in SQ. 
The authors are indebted to V. Charmandaris for having provided his deep near-IR images of Stephan's Quintet taken with the \textit{WIRC} instrument on the Palomar 200-inches telescope.
We are grateful to M.G. Allen for making publicly available the \textit{Mappings III} shock library. We also thanks the anonymous referee for useful comments that helped us improve the clarity of the manuscript and for having suggested the use of optical data.
\end{acknowledgements}

\bibliographystyle{aa.bst}
\bibliography{SQ_Dust.bbl}

\Online

\begin{appendix}

\section{Fitting PAH features}
\label{appendix_PAHFit}

\begin{figure*}
    \centering
     \includegraphics[width = 0.7\textwidth]{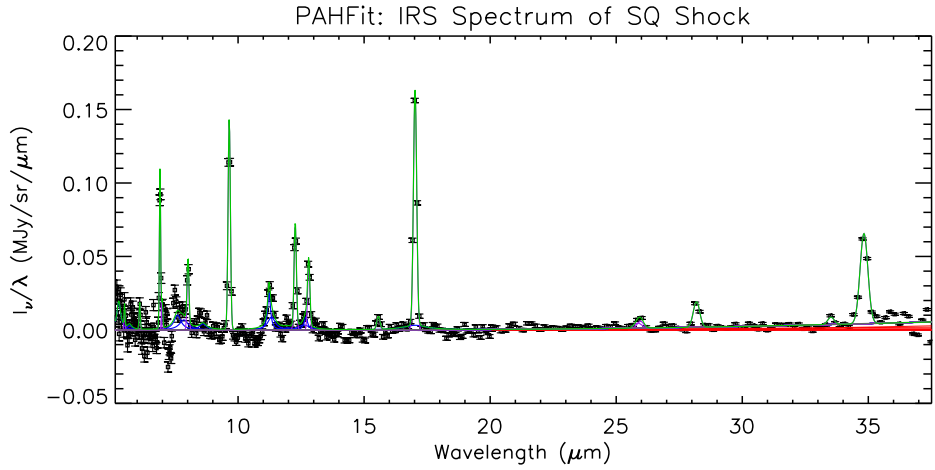}
 \caption{Result of a PAHFit run on the $5-38\,\mu$m \textit{Spitzer IRS} spectrum extracted over a $\sim 18" \times 15"$ area centered on the position in the center of the ridge (ON beam on Fig.~\ref{Fig_NUV_16_24_70mic_contours}). The green line shows the complete model. The fitted gas lines are in magenta. The gaussian profiles of the gas lines are used to remove their contributions and extract a ``pure dust spectrum'' (see Fig.~\ref{Fig_pahfit_noext_PAH_SQ_smooth}).
}
     \label{Fig_pahfit_noext_full_SQ}
  \end{figure*}

\begin{figure*}
    \centering
     \includegraphics[width = 0.7\textwidth]{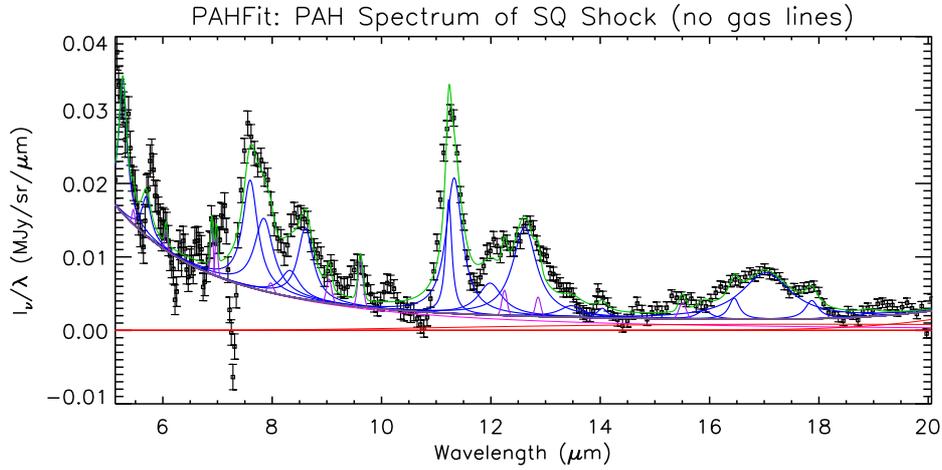}
 \caption{PAHFit decomposition of the \textit{Spitzer IRS} dust spectrum (smoothed over 5 resolution elements, and from which gas lines have been removed) extracted in the center of the SQ ridge. Blue solid lines shows the lorentzian components of the PAH decomposition, and the thick gray line is the total (dust + starlight) continuum. The result of the fit is the green solid line. 
}
     \label{Fig_pahfit_noext_PAH_SQ_smooth}
  \end{figure*}

\begin{figure*}
    \centering
     \includegraphics[width = 0.7\textwidth]{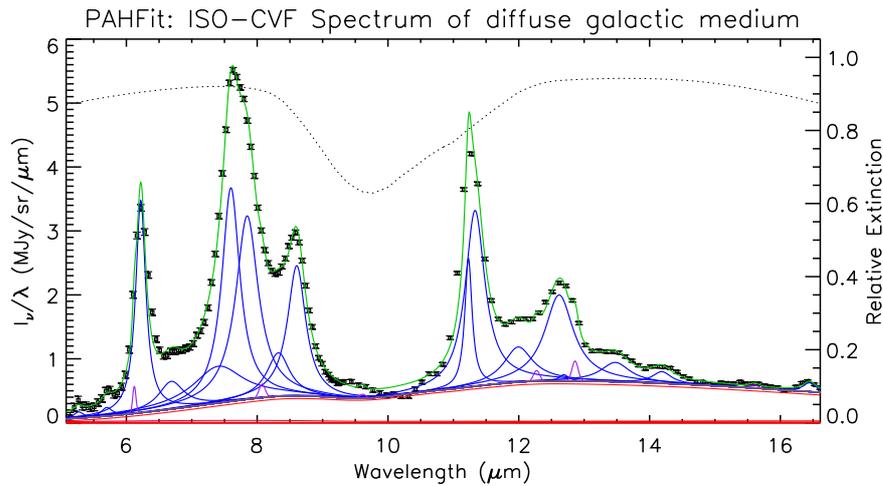}
 \caption{PAHFit decomposition of the \textit{ISO-CVF} spectrum of the diffuse Galactic light \citep{Flagey2006}, centered on the Galactic coordinates $(26.8, +0.8)$. The gas lines have been removed from the spectrum. The lorentzian components of the decomposition of the PAH features are shown in blue. All components are diminished by the extinction, indicated by the dotted black
line, with axis at right. The solid green line is the full fitted model, plotted on the observed flux intensities and uncertainties. 
}
     \label{Fig_pahfit_ext_CVF}
  \end{figure*}

The results of the PAHFit decompositions of the \textit{Spitzer IRS} spectrum of the center of SQ ridge and, for comparison, of the \textit{ISO-CVF} spectrum of the diffuse Galactic medium, are presented in Fig. ~\ref{Fig_pahfit_noext_full_SQ}, \ref{Fig_pahfit_noext_PAH_SQ_smooth} and \ref{Fig_pahfit_ext_CVF}. No extinction parameter is introduced in the fit. 


\end{appendix}

\end{document}